\documentclass[journal=jacsat,manuscript=article]{achemso}
\usepackage[version=3]{mhchem}
\newcommand{\angstrom}{\mbox{\normalfont\AA}}
\usepackage{graphicx}
\usepackage{xcolor}
\usepackage{pgfplots}
\usepackage{hyperref}
\usepackage{tikz}
\usepackage{subfigure}
\usetikzlibrary{shapes,arrows,chains,fit,calc,shapes,matrix,positioning,patterns}
\pgfplotsset{compat=newest}
\usepgfplotslibrary{groupplots}
\usepackage{comment}
\setkeys{acs}{keywords = true}
\usetikzlibrary{external}

\author{Lorenzo Tinacci}
\affiliation[Università degli Studi di Torino]{Dipartimento di Chimica, via P. Giuria 7, 10125 Torino, Italy}
\alsoaffiliation[Université Grenoble Alpes]{Institut de Planétologie et d’Astrophysique de Grenoble (IPAG), 38000 Grenoble, France}
\author{Aurèle Germain}
\affiliation[Università degli Studi di Torino]{Dipartimento di Chimica, via P. Giuria 7, 10125 Torino, Italy}
\author{Stefano Pantaleone}
\affiliation[Università degli Studi di Torino]{Dipartimento di Chimica, via P. Giuria 7, 10125 Torino, Italy}
\alsoaffiliation[Università di Perugia]{Dipartimento di Chimica, Biologia e Biotecnologie, 06123 Perugia, Italy}
\author{Stefano Ferrero}
\affiliation[Universitat Autonoma de Barcelona]{Departament de Quimica, 08193 Bellaterra, Catalonia, Spain}
\affiliation[Università degli Studi di Torino]{Dipartimento di Chimica, via P. Giuria 7, 10125 Torino, Italy}
\author{Cecilia Ceccarelli}
\affiliation[Université Grenoble Alpes]{Institut de Planétologie et d’Astrophysique de Grenoble (IPAG), 38000 Grenoble, France}
\author{Piero Ugliengo}
\affiliation[Università degli Studi di Torino]{Dipartimento di Chimica, via P. Giuria 7, 10125 Torino, Italy}
\email{piero.ugliengo@unito.it}

\title{Theoretical distribution of the ammonia binding energy at interstellar icy grains: a new computational framework}

\abbreviations{BE,ISM}
\keywords{Amorphous water ice, xTB-GFN2, ONIOM, DLPNO, B97D3, NH$_3$ adsorption, NH$_3$ binding energy}

\begin{document}


\begin{tocentry}
\includegraphics{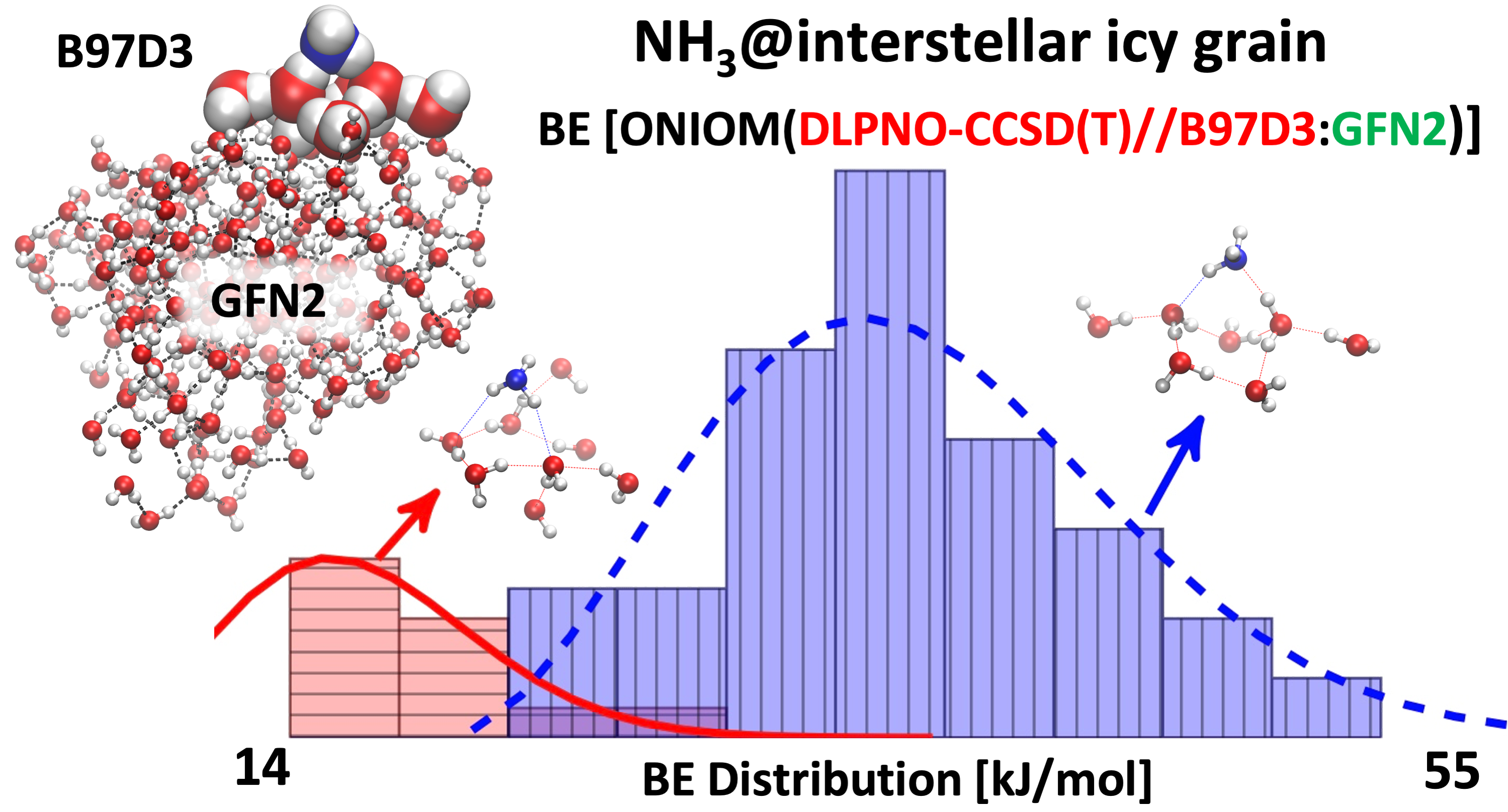} \label{For Table of Contents Only}
\end{tocentry}

\begin{abstract}
The binding energies (BE) of molecules on the interstellar grains are crucial in the chemical evolution of the interstellar medium (ISM). 
Both temperature programmed desorption (TPD) laboratory experiments and quantum chemistry computations have often provided, so far, only single values of the BE for each molecule. 
This is a severe limitation, as the ices enveloping the grain mantles are structurally amorphous, giving rise to a manifold of possible adsorption sites, each with different BEs. 
However, the ice amorphous nature prevents the knowledge of structural details, hindering the development of a common accepted atomistic icy model. 
In this work, we propose a computational framework that closely mimics the formation of the interstellar grain mantle through a water by water accretion. 
On that grain, an unbiased random (but well reproducible) positioning of the studied molecule is then carried out. 
Here we present the test case of NH$_3$, an ubiquitous species in the molecular ISM. 
We provide the BE distribution computed by a hierarchy approach, using the semiempirical xTB-GFN2 as low-level method to describe the whole icy cluster combined with the B97D3 DFT functional as high-level method on the local zone of the NH$_3$ interaction. 
The final ZPE corrected BE is computed at ONIOM(DLPNO-CCSD(T)//B97D3:xTB-GFN2) level, ensuring the best  cost/accuracy ratio. 
The main peak of the predicted NH$_3$ BE distribution is in agreement with experimental TPD and literature computed data.
A second broad peak at very low BE values is also present, never detected before.
It may provide the solution to a long-standing puzzle about the presence of gaseous NH$_3$ observed also in cold ISM objects.
\end{abstract}

\section{Introduction}

Interstellar dust grains in cold ($\sim 10$ K) molecular clouds are made up of submicro-meter sized refractory cores (mainly silicates and carbonaceous material) on top of which water molecules are formed \textit{in situ} through reactions involving hydrogen and oxygen \cite{tielens1982model,molpeceres_2019,dulieu2010experimental,jing2011water,oba2012water}. 
Eventually, this process leads to the accretion of a thick (made up of more than 100 layers: \textit{e.g.} Taquet et al. 2012 \cite{taquet2012multilayer}) amorphous icy mantle.  
At the same time, other atoms and molecules formed in the gas-phase can condensate and be adsorbed onto the grain mantles, where they may diffuse and react on the icy surfaces enriching the chemical composition of the grain mantles. 

The vast majority of the species frozen or trapped on the grain mantles are only observable when they are released into the gas-phase either in warm ($\geq 100$ K) regions, such as hot cores/corinos via thermal desorption \cite{blake1987molecular,charnley1992molecular,ceccarelli2000hot}, or in shocked regions, via sputtering of the mantles \cite{bachiller1993high,flower1994grain,lefloch2017l1157}.
All the processes mentioned above, adsorption and diffusion, as well as desorption, are governed by a key parameter, the so-called Binding Energy (BE), namely the strength of a species to remain glued to the surface.
Since BE has exponential dependence into the expressions of astrochemical models that describe the above processes, their estimation with a good level of accuracy is crucial for our knowledge of their chemical evolution\cite{penteado2017sensitivity}. This fundamental piece of information can be obtained either via theoretical or experimental approaches.

Usually, estimates of PES via theoretical methods involve computations using a molecular mechanics force-fields approach and/or rigorous quantum mechanical methods. 
In both  cases, an atomistic model of the icy grain is needed and the BE is computed via a super-molecular approach, namely computing the difference between the energy of the adsorbate interacting with the icy grain and the energies of the free adsorbate and the original icy grain. 
Despite this simple definition, the final BE value can be affected by many factors, both from modelling and methodological points of view. 
To start with, the computerized icy model is usually ill-defined, as the structure of the interstellar ice is poorly known. 
Therefore, a variety of models to simulate the ice-species adsorption has been proposed in the literature, from just a single water molecule up to periodic models of either crystalline or amorphous water ice \cite{wakelam2017binding,das2018approach,ferrero2020binding}.
Due to the difficulty of simulating the icy grain accretion by \textit{in situ} water formation, all the models so far are constructed by assembling a variety of already formed water molecules interacting through hydrogen bonds. \cite{shimonishi2018adsorption,rimola2018can}
This may have serious consequences on the final ice structure, as the fraction of water formation energy transferred to the grain can affect its final structural features much more than the mere hydrogen bond interaction between the water molecules \cite{Pantaleone_2021}.
In addition, it has theoretically and experimentally been shown that any species does not have a single BE on amorphous water surfaces (AWS) but rather a distribution of BE, which depends on the species and the surface \cite{amiaud2006interaction,ferrero2020binding, bovolenta2020high,Molpeceres2020-binding, he2016binding}.
Therefore, the icy grains should be large and varied enough to allow to reconstruct the BE distribution of a species and not just a value.
To overcome the above mentioned problems, we have recently proposed \cite{Aurele_grain} an automatic and unbiased approach to construct water ice clusters and obtain the binding energy distribution of any species (see the Methodology section for further details).

Beside the icy model definition, the second important issue to compute the BE is the adopted level of theory, which always represents compromise between the computation accuracy and the computational cost (method and system size).

Methods based on the molecular mechanics may reach some accuracy when designed to treat very specific cases but fail for cases outside their specific parametrization. Alternatively, methods based on the best level of quantum chemistry, like the golden standard CCSD(T)\cite{ccsd(t)}, ensure a well balanced treatment of all the relevant interactions responsible for the adsorption on the icy grain surface, irrespective on the considered adsorbate molecule.
However, the computational time required by CCSD(T) grows too steeply to be applicable to large icy grains. 

Here, we propose a new method which optimises the computation accuracy on very large icy grain models.
Specifically, we implemented an automatic procedure which is based on the ACO-FROST code, recently developed by our group \cite{Aurele_grain}, to construct a large ($\geq 1000$ water molecules) icy grain.
Briefly, only a selected portion of the icy grain, where the adsorption takes place, is treated at a very high level of theory, while the whole cluster is treated at a lower level. 
This procedure itself is not completely new, as it has been already adopted in the field of surface science adsorption \cite{sauer} and for some ice models \cite{song2016formation,molpeceres2020adsorption,sameera2017oniom,sameera2020ch3o,ferrero2020binding}. 
In a recent work, we adopted a similar scheme to improve the BEs computed for a set of molecules on periodic ice models (both crystalline and amorphous) reaching a CCSD(T) quality results \cite{ferrero2020binding}.
Similarly, Duflot et al\cite{toubin2021ONIOM} adopted a QM:MM approach using for the QM method the DLPNO-CCSD(T) technique\cite{DLPNO_CCSD(T)}, a very accurate and computational feasible version of the CCSD(T) standard based on localized orbitals and the PM6 Semi-empirical method\cite{PM6} for the rest of the system. 

Our newly proposed procedure, described in this work, possesses the following novelties with respect to the above works:
\begin{enumerate}
\item [i)] an unbiased procedure to generate a large variety of adsorbed structures, not dependent on the nature of the adsorbate molecule and the size of the icy cluster, which allows to compute a BE distribution of the considered species;
\item [ii)] the low-level theory adopted to treat the whole icy cluster is based on the accurate semi-empirical tight-binding xTB-GFN2 method, very recently developed by the Grimme's group \cite{GFN2};
\item [iii)] the high-level theory adopted to describe the ice around the adsorbing site is based on the DLPNO-CCSD(T) method with a selection of large Gaussian basis sets.
\end{enumerate}
In addition, the procedure is carried out automatically by a package of Python scripts, which allow the the construction, submission and data extraction of the needed calculations.

The BE values resulted from the above approach should in principle be compared with experimental derivations of BEs. 
However, this is not straightforward for the following reasons.
Binding energies are usually experimentally derived via the so-called Temperature Programmed Desorption (TPD) method.
Strictly speaking, this method provides the Desorption Activation Energy (DAE), which is often interpreted as BE. 
In practice, the DAE is derived indirectly from the TPD peaks through the Readhead’s method \cite{redhead}, or more sophisticated numerical techniques. 
In most TPD experiments, a water ice surface hosts a mono-layer of the adsorbate and, therefore, the BE depends also on the surface coverage \cite{he2016binding}. 
This renders the comparison between DAE and the computed BE actually not straightforward \cite{king}. 
For example, ice restructuring processes may affect the ﬁnal DAE, making it different form the BE. 
Also, sometimes TPD experiments only provide desorption temperature peaks T$_{des}$, with no numerically estimate of the DAE. 
For instance, Collings et al \cite{collings2004laboratory} computed the BE of a X species as: BE(X) = [T$_{des}$(X)/T$_{des}$(H$_2$O)] BE(H$_2$O), in which T$_{des}$(X) is the desorption temperature of the X species in constrast with that of water T$_{des}$(H$_2$O), by assuming BE(H$_2$O) = 4800 K ($\sim 40$ kJ/mol). 
For the above reasons, a one-by-one comparison between experiment and modeling should be carried out with extreme care, particularly when a BE distribution is computed, as in the present work.

For our first application of the new method presented here, we chose the ammonia molecule, because it is a very studied and important species in the molecular ISM. 
It is the first detected interstellar polyatomic molecule \cite{cheung1968detection}, and one of the most observed, ubiquitous and studied.
It is found in the gaseous form towards the Galactic Center warm molecular clouds and cores \cite{cheung1968detection,winnewisser1979ammonia}, diffuse clouds \cite{liszt2006comparative}, massive hot cores \cite{morris1973interstellar}, molecular outflows \cite{umemoto1999ortho}, solar-type protostars \cite{mundy1990circumstellar}, cold molecular clouds \cite{irvine1987chemical}, prestellar cores \cite{crapsi2007observing} and protoplanetary disks \cite{salinas2016first}.
Ammonia is also observed to be very abundant in the icy mantles that envelope the interstellar dust grains in cold regions \cite{knacke1982observation}.
Obviously, whether ammonia is either in the gaseous or solid form is governed by its BE.
On the same vein, understanding the ammonia chemistry requires to have a good knowledge of the ammonia BE and, more specifically, its BE distribution, which is the focus of this work.

\section{Methodology}\label{sec:Methodology}

\subsection{Icy grain model and NH$_3$ binding site sampling}\label{subsec:ACO_frost}

The water-ice grain model used throughout this work, the binding energy sampling procedure and the preliminary BE geometric optimization were taken from a previous work by our group, which is summarized in this section.\cite{Aurele_grain}

\paragraph{Water-ice grain model.}
In order to build-up the grain model a bottom-up approach was followed, \textit{i.e.} by random successive aggregations of water molecules. 
A geometry optimization was performed at each addition of water molecule, followed by a short molecular dynamics (MD) run at 10 K every 10 added H$_2$O molecule, to mimic the induced thermal motion due to the partly transferred energy of water formation\cite{Pantaleone_2021} occurring in the real grain, but not taken into account here (\textit{vide supra}).

As already discussed in the Introduction, our grain model includes 200 water molecules, large enough to allow for a proper sampling of many adsorbing sites compared to previously adopted models. 
The grain construction was performed at a semi-empirical level using the xTB (v.6.3.3) \cite{xtb} code (GFN2\cite{GFN2} and the force field GFN-FF methods\cite{Spicher2020}) developed by the Grimme's group at the Bonn University.

\paragraph{Binding energy sampling site procedure.}
The NH$_3$ binding site sampling was done by placing a grid consisting of 12 vertexes (forming an icosahedron), which were tightened for a total of 162 vertexes uniformly spread around the grain \cite{meng2019universal}. 
The grid points were projected closer to the grain surface and each point was substituted by a randomly oriented ammonia molecule with respect to the direction vector joining the N atom and the grain center of mass. 
The projection brings a distance between 2.5-3 \angstrom $\,$ from the grain, used to positioning $\mathrm{NH_3}$.

\begin{figure}
\centering
\includegraphics[width=0.99\textwidth]{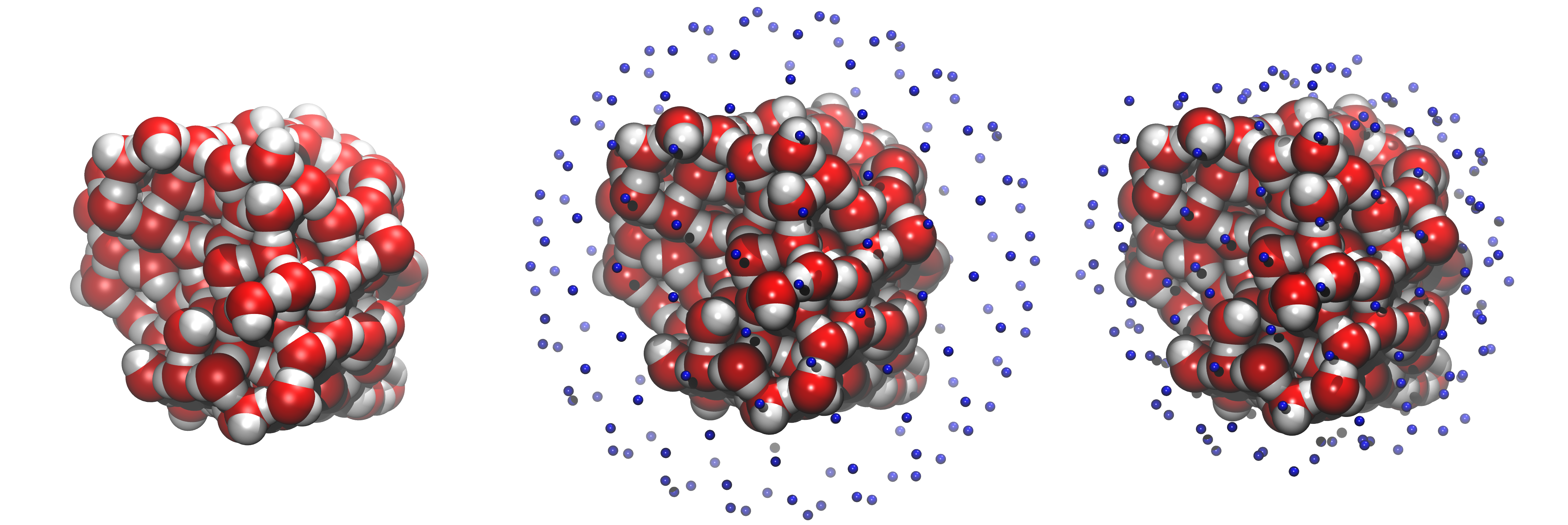}
\caption{(Left) Icy grain model, (Central) with the overlapped 162 vertices grid points in blue color and (Right) with the same vertices projected closer (2.5-3 \angstrom) to the grain surface. Atom color legend: oxygen in red, hydrogen in white. Data taken from Ref.\cite{Aurele_grain}.}
\label{fig:grain_grid}
\end{figure}

\paragraph{Preliminary geometry optimization.}
After the NH$_3$ sampling, a preliminary geometry optimization via xTB-GFN2\cite{GFN2,xtb} was performed. 
Two subsequent geometry optimizations were carried out in which: i) only the NH$_3$ molecule was set free to relax on the grain, while all the water molecules were kept fixed at the optimized free grain positions; and ii) the atomic positions of NH$_3$ and the water molecules included within a cutoff distance of $5\,\angstrom$ from the NH$_3$ were relaxed, while the remaining water molecules were kept fixed. 
This choice enforces the structural rigidity experienced by the water molecules in a real (and much larger) icy grain. 
During the second task, we found cases where the number of the mobile water molecules changed during the optimization procedure, due to the rearrangements of both the NH$_3$ and the water molecules within the selected zone. 
In these cases, the described cycle was repeated by selecting again a new mobile zone, and re-optimize the structure until no changes in the number of water molecules occurred.  

\subsection{Computational Methods}
After a preliminary geometry optimization with the xTB (v.6.3.3) \cite{xtb} computational program, the refined binding energy distribution of ammonia on the amorphous ice model was obtained by combing the tools implemented in three codes: xTB (v.6.3.3), Gaussian (v.16, Revision B.01)\cite{g16}, and ORCA (v.4.2.1) \cite{ORCA}. 
We relied on the multilevel ONIOM\cite{ONIOM_gau}(DFT:xTB-GFN2) approach as implemented in the Gaussian program to obtained accurate optimized geometries. 
As the GFN2\cite{GFN2} method has not been implemented in the Gaussian program yet, xTB (v.6.3.3)\cite{xtb} was called as external program to work on the low level zone of the ONIOM method. 
Finally, the energies of the high level zone where refined with ORCA (v.4.2.1) \cite{ORCA} at DLPNO-CCSD(T)\cite{DLPNO_CCSD(T)_new} level of theory. 
Rendering of molecule images have been obtained via the VMD software\cite{vmd}, while the graphics elaboration and plots via TikZ and PGFPlots \LaTeX $\,$ packages.

\paragraph{ONIOM method.}
The ONIOM ('Our own N-layered Integrated molecular Orbital and Molecular mechanics') method\cite{ONIOM} is a hybrid approach that enables different \textit{ab initio}, semi-empirical or classical mechanics-based methods to be combined to different parts of a system to give reliable geometry and energy at reduced computational cost.
All the calculations were performed with the two-layer ONIOM(QM:SQM) method. In the specific: the zone, in which the quantum-mechanical method (QM) is used (also called Model zone), consists of NH$_3$ and neighboring water molecules within 5 \angstrom $\,$ from NH$_3$, while the whole system (Real zone) is treated at semi-empirical  quantum mechanical  (SQM) level.  
The total energy (E), gradient vector ($\mathcal{G}$) and Hessian matrix ($\mathcal{H}$) for the ONIOM(QM:SQM) two-layer set up are, therefore:
\begin{subequations}
\begin{align}
    \mathrm{E}  &= \mathrm{E(R{:}SQM)} + \mathrm{E(M{:}QM)} - \mathrm{E(M{:}SQM)} \; , \label{eq:E_oniom}\\
    \mathcal{G} &= \mathcal{G}(\mathrm{R{:}SQM}) \times \mathcal{J_\mathrm{M:R}} + \mathcal{G}(\mathrm{M{:}QM}) \times \mathcal{J_\mathrm{M:R}} - \mathcal{G}(\mathrm{M{:}SQM}) \times \mathcal{J_\mathrm{M:R}} \; ,\label{eq:G_oniom}\\
    \mathcal{H} &= \mathcal{H}(\mathrm{\mathrm{R{:}SQM}}) \times \mathcal{J_\mathrm{M:R}} +  \mathcal{J^\mathrm{T}_\mathrm{M:R}} \times \mathcal{G}(\mathrm{M{:}QM}) \times \mathcal{J_\mathrm{M:R}} -  \mathcal{J^\mathrm{T}_\mathrm{M:R}} \times \mathcal{G}(\mathrm{M{:}SQM}) \times \mathcal{J_\mathrm{M:R}} \; ,
\end{align}
\end{subequations}
where $\mathcal{J_\mathrm{M:R}}$ is the Jacobian matrix between the Model (M) and the Real (R) nuclei.

The binding energy (BE, positive for a bounded system), is defined as the opposite of the interaction energy, the last quantity being the difference  between the energy of the complex between the grain and the adsorbate ($\mathrm{E}_{c}$) and the sum of the energies of the isolated adsorbate ($\mathrm{E}^{iso}_{ads}$) and the isolated grain ($\mathrm{E}^{iso}_{grn}$). 
The equation adopted for the calculation of the ONIOM BEs, after equation \ref{eq:E_oniom}, are:
\begin{equation}\label{eq:BE}
\mathrm{BE} = - \Delta \mathrm{E} = \mathrm{E}^{iso}_{ads}(\mathrm{QM}) + \mathrm{E}^{iso}_{grn}(\mathrm{QM{:}SQM}) - \mathrm{E}_{c}(\mathrm{QM{:}SQM}) \; ,
\end{equation}
where the energies of the isolated systems are referred to the specified level at which geometry are also optimized.
BEs can be decomposed in the pure electronic interaction ($\mathrm{BE}_e$) corrected for the Basis Set Superposition Error (BSSE) and the deformation energy ($\mathrm{\delta E}_{def}$) contributions. 

The $\mathrm{BE}_e$ is given by:
\begin{equation}\label{eq:BSSE}
    \mathrm{BE}_{e} = \mathrm{E}^{iso//c}_{ads}\bigl(\mathrm{\mathcal{G}(grn)\bigr)}+ 
    \mathrm{E}^{iso//c}_{grn}\bigl(\mathrm{\mathcal{G}(ads)\bigr)} -\mathrm{E}_{c}(\mathrm{QM}) \; ,
\end{equation}
where $\mathrm{E}^{iso//c}_{ads}\bigl(\mathrm{\mathcal{G}(grn)\bigr)}$ and $\mathrm{E}^{iso//c}_{grn}\bigl(\mathrm{\mathcal{G}(ads)\bigr)}$ are the energies of the isolated adsorbate and the grain in the geometries assumed in the complex ($iso//c$) in presence of the ghost orbitals of the grain $\mathrm{\mathcal{G}(grn)}$ and the adsorbate $\mathrm{\mathcal{G}(ads)}$, respectively. 
Obviously, as the BSSE is already taken into account by the definition in the GFN2 method, equation \ref{eq:BSSE} only applies to the QM methods (\emph{vide infra}) on the Model zone.

The $\mathrm{\delta E}_{def}$ is given by:
\begin{equation}
    \mathrm{\delta E}_{def} = \underbrace{\bigl( \mathrm{E}^{iso//c}_{ads}-\mathrm{E}^{iso}_{ads}\bigr)}_{\mathrm{\delta E}^{ads}_{def}} +\underbrace{\bigl( \mathrm{E}^{iso//c}_{grn}-\mathrm{E}^{iso}_{grn}\bigl)}_{\mathrm{\delta E}^{grn}_{def}} \; ,
\end{equation}
where $\mathrm{\delta E}^{ads}_{def}$ and $\mathrm{\delta E}^{grn}_{def}$ are the deformation energy of the adsorbate and the surface, respectively.
Obviously, $\mathrm{\delta E}_{def}$ is always a positive quantity.

Moreover, vibrational frequencies were computed on the Model zone to obtain the zero-point energies (ZPE), from which the $\Delta$ZPE resulted as:
\begin{equation}
    \Delta \mathrm{ZPE} = \mathrm{ZPE}_{c} -  \mathrm{ZPE}^{iso}_{ads} -  \mathrm{ZPE}^{iso}_{grn} \; .
\end{equation}

Including all the above-mentioned contributions, Equation \ref{eq:BE} becomes:
\begin{equation}
    \label{eq:BE_decompose}
    \mathrm{BH(0)} = \underbrace{\mathrm{BE}_e - \bigl(\mathrm{\delta E}^{grn}_{def} + \mathrm{\delta E}^{ads}_{def}\bigr)}_{\mathrm{BE}} - \Delta \mathrm{ZPE} \; .
\end{equation}

In our ONIOM setup, the low-level layer was treated with the xTB-GFN2 semi-empirical quantum mechanical (SQM) method \cite{GFN2}, working as an external program with Gaussian16. 
The default xTB-GFN2 parameters were used for the SCF. 
On the high-level layer two different methods were used in order to compute subsequent tasks: 
\begin{itemize}
\item \textbf{Geometry optimization and frequency calculations} -- The B97D3\cite{B97D,DFT-D3} functional, as implemented in Gaussian16, with the aug-cc-pVTZ basis set\cite{kendall1992dunning} and the default setup for geometry optimization, SCF and integral grid density. 
\item \textbf{Final energy refinement} -- DLPNO-CCSD(T) method,\cite{DLPNO_CCSD(T),DLPNO_CCSD(T)_new} as implemented in ORCA,  with the aug-cc-pVTZ as primary basis set, while the aug-cc-pVTZ/C\cite{pvtz_c} as auxiliary basis set for the resolution of the identity (RI) approximation in electron repulsion integrals. 
All these calculations were carried out with a tight-PNO set up and the default settings for the SCF.
\end{itemize}
During the ONIOM geometry optimization all atoms outside the Model zone were kept fixed; only mechanical embedding and no micro-iterations were used. 
In the frequency calculations (calculated in the harmonic approximation), only the normal modes related to the nuclei inside the Model zone were taken into account, keeping fixed all the other nuclei.

The treatment of the isolated icy surface required extra care, as the Model zone may change during the search for the optimum structure when the grain is adsorbing the NH$_3$ molecule. 
Therefore, to ensure a proper coherence, we used in this section, as the Model zone for evaluating the energy $\mathrm{E}^{iso}_{grn}$ of the free grain, the very last set of water molecules defined in the cycling procedure described above, on an otherwise unique and fixed reference geometry of the free cluster.

\begin{figure}
\centering
\includegraphics[width=0.31\textwidth]{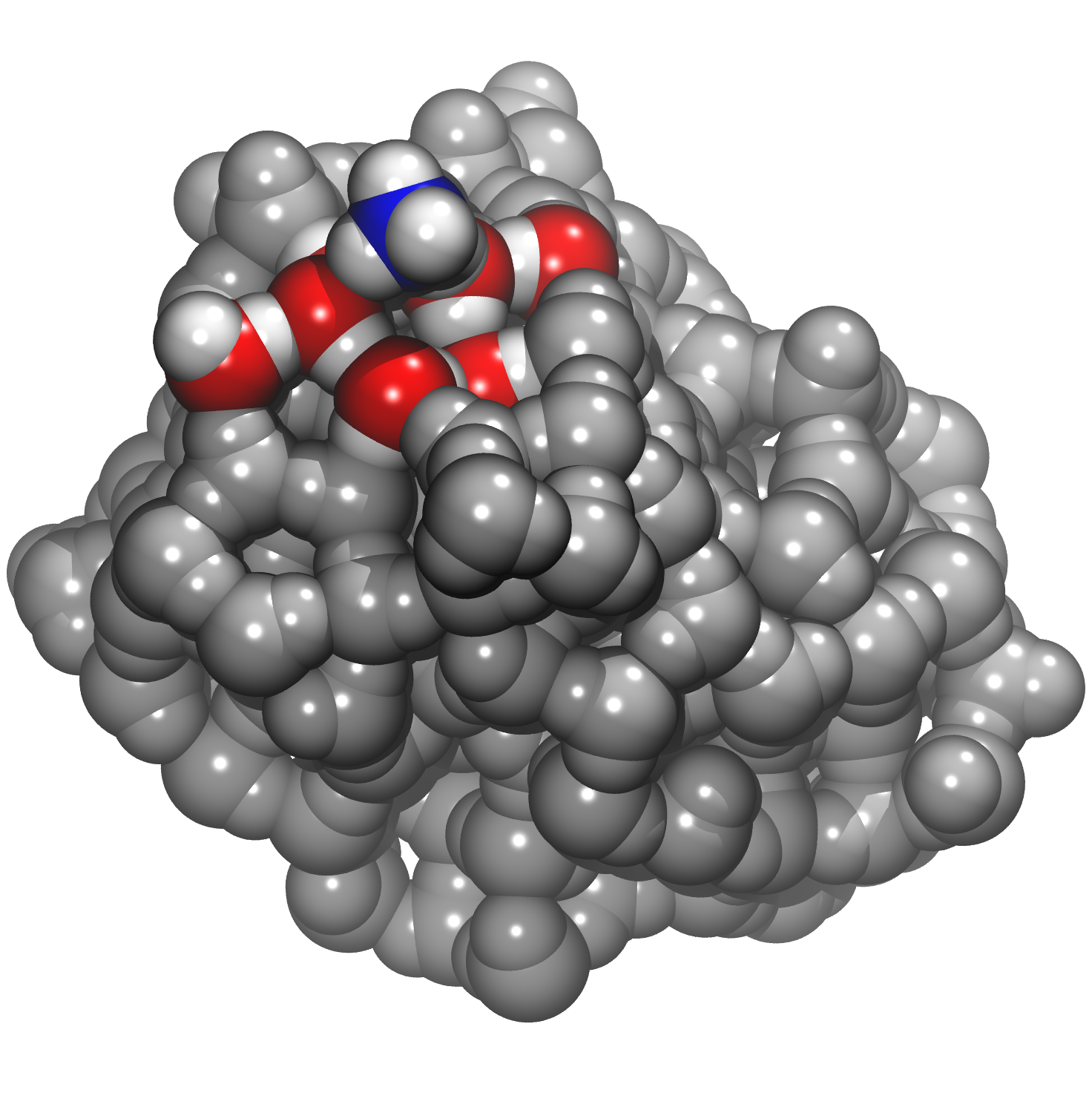}
\hfill
\includegraphics[width=0.31\textwidth]{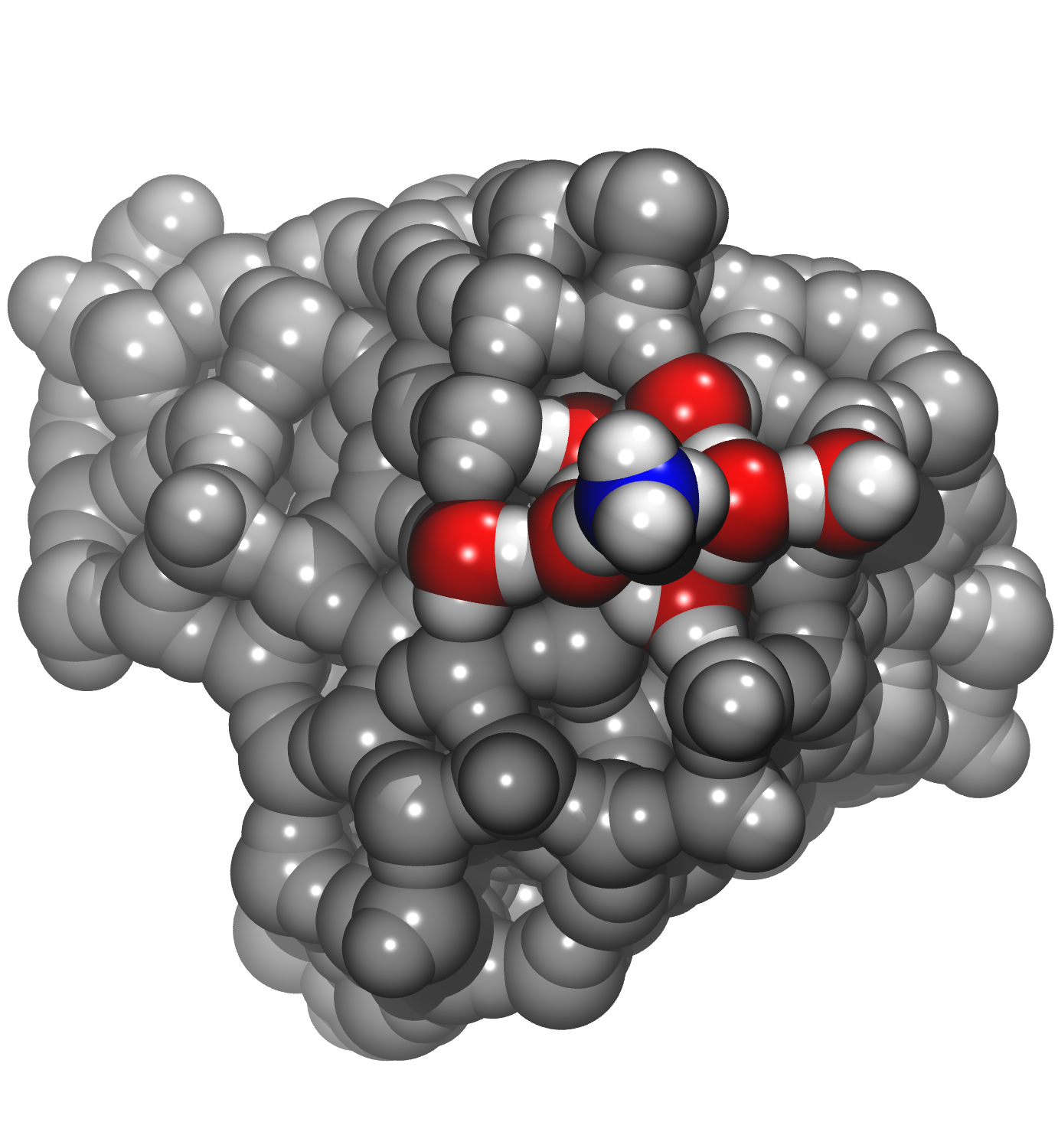}
\hfill
\includegraphics[width=0.31\textwidth]{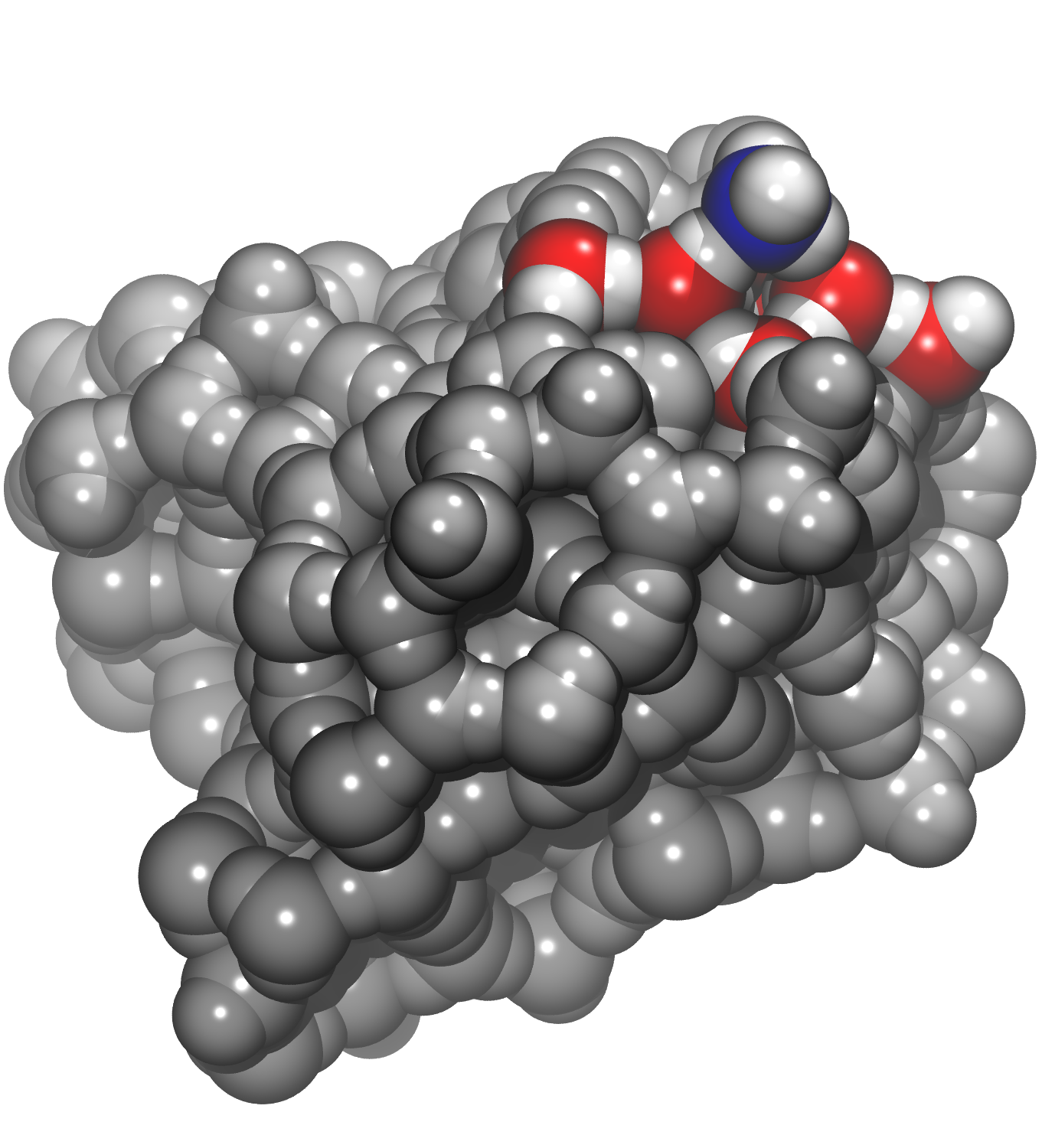}
\caption{Three different perspectives showing the ONIOM zones: the atoms in the Model (high-level) zone are in colors while the low-level zone of the system is pictured in grey color. Atom color legend: oxygen in red, nitrogen in blue, hydrogen in white}
\label{fig:my_label}
\end{figure}

\subsection{Model zone setup}
The definition of the Model zone, which is the core of any ONIOM-based procedure, implies to choose the proper level of theory but also the number of water molecules to be included in the QM description.

\paragraph{Geometry optimization constraints.}
We adopted the same strategy for the ONIOM calculation used for the optimizations performed with the GFN2\cite{GFN2} level. 
However, since the method to treat the Model zone is computational demanding, a less tight criterion on the optimization convergence was used: when the number of water molecules of the Model zone changes by $\geq|2|$ units in the Model zone, we run further geometry optimizations with the redefined model Model zone, until the above condition is satisfied.

\paragraph{Model zone size benchmark.}
The Model zone defined within $5\,\angstrom$ from the NH$_3$ relies on a trade-off between two main requirements: i) including all the local NH$_3$-H$_2$O interactions; and ii) saving computational resources. 

In order to understand the influence of the Model zone size on the BE, a benchmark was performed taking the single point energy evaluation of 8 different optimized cases with the standard Model zone definition ($5\,\angstrom$) and expanding its size from 5 up to $8.5\,\angstrom$ (which corresponds to include up to 21--34 water molecules) while keeping the geometry of the whole system fixed. 

\begin{figure}
\begin{center}
\includegraphics[width=0.6\columnwidth]{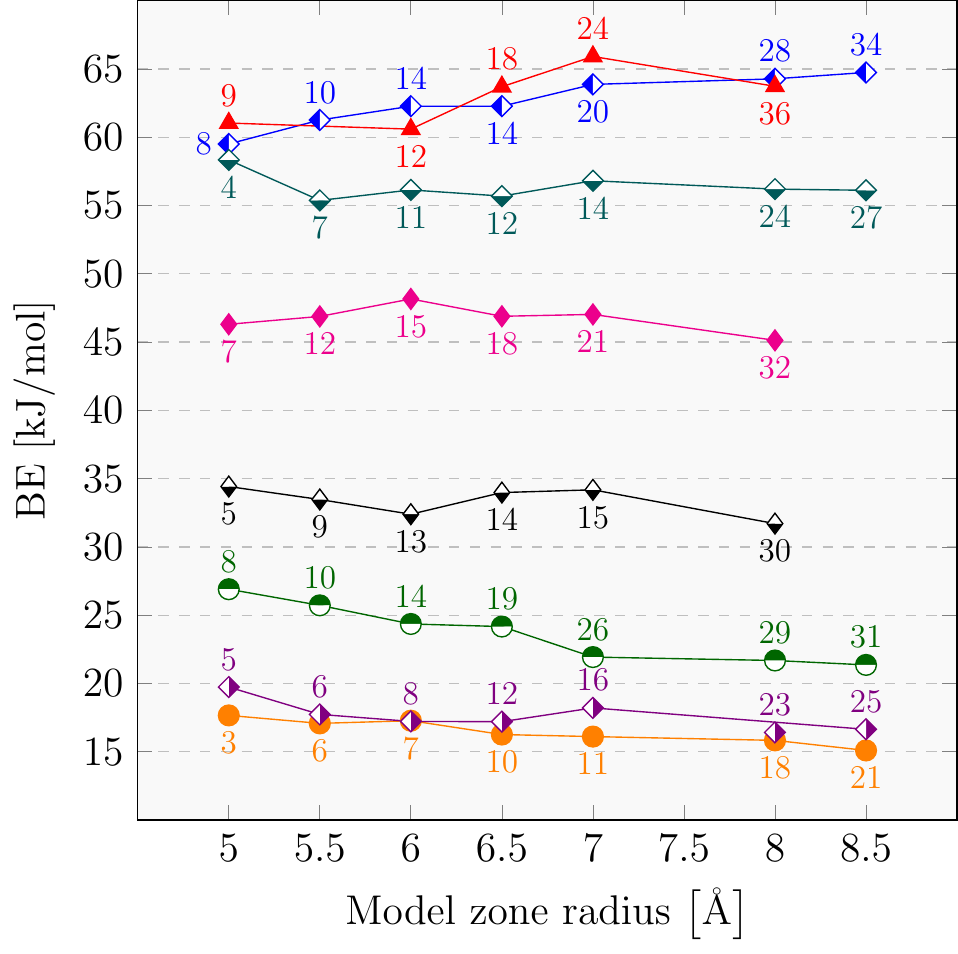}
\end{center}
\caption{BSSE corrected BEs calculated at ONIOM(B97D3/aug-cc-pTVZ:xTB-GFN2) level as function of the Model zone size. Each symbol/color represents the same BE sample while the number of water molecules inside the Model zone is reported close the related symbol.}
\label{fig:be_radius}
\end{figure}

Single point energy calculations were carried out at the same level of theory described in the previous section, \textit{i.e.} ONIOM(B97D3/aug-cc-pVTZ:xTB-GFN2). 
Figure \ref{fig:be_radius} shows for all but two samples, a change in the the BE value well within 5 kJ/mol and a rather flat variation in the BE values. 
The two exceptions are at the limit of the threshold of 5 kJ/mol,  (\textit{i.e.} within the chemical accuracy limit).

\begin{figure}
\begin{center}
\includegraphics[width=0.5\columnwidth]{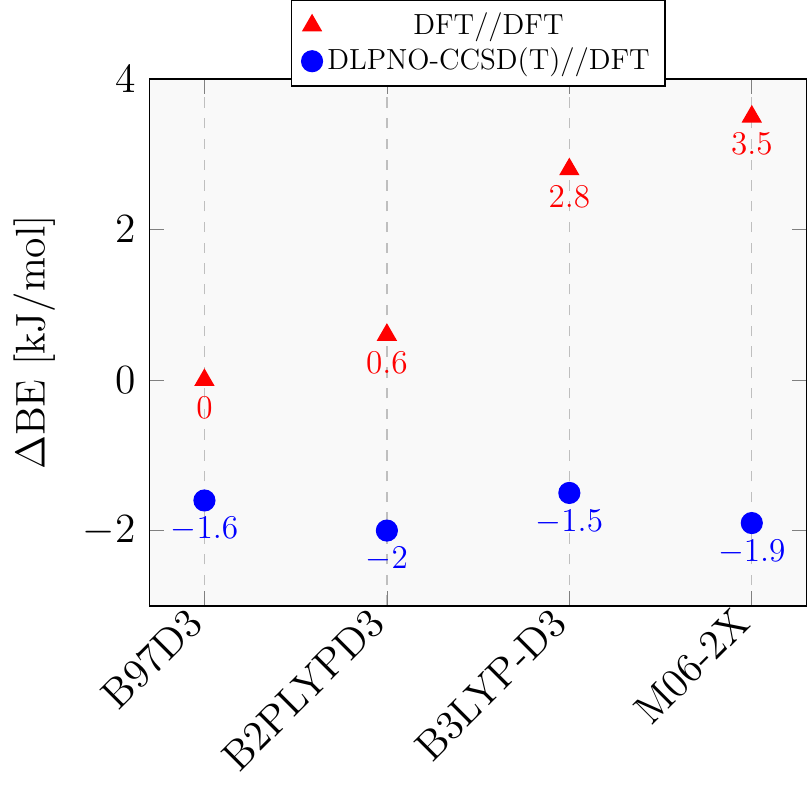}
\end{center}
\caption{Differences $\Delta$BE between the BE B97D3 reference value (48.4 kJ/mol) and the BEs computed with the reported QM methods, all coupled with a aug-cc-pVTZ basis set quality and BSSE corrected.}
\label{fig:bench_meth}
\end{figure}

\paragraph{Model zone methods benchmark.}
The pure GGA B97D3 functional\cite{B97D,DFT-D3} used to deal with the Model zone is well apt to deal with non-covalent interactions like the one responsible of the grain cohesion and the NH$_3$ BE\cite{B97D}.
To assess the B97D3 performance for the present case,  we compared, for one selected NH$_3$/grain case, structures and BEs (corrected for BSSE) with: i) the B2PLYPD3 double-hybrid functional with empirical dispersion corrections\cite{B2PLYPD3}; ii) the B3LYP\cite{B3LYP_0,B3LYP_1} with D3 version of Grimme’s dispersion with Becke-Johnson damping function\cite{DFT-D3}; and iii) the Minnesota double-exchange M06-2X functional\cite{M06-2X}, coupled with the aug-cc-pVTZ\cite{kendall1992dunning} basis set. 
DFT BEs were then refined at DLPNO-CCSD(T)/(aug-cc-pVTZ \& aug-cc-pVTZ/C) tight-PNO level (all the values corrected for the BSSE) computed at each DFT geometry optimum. 
The results are presented in figure \ref{fig:bench_meth}. 
Among all adopted functionals, the B97D3 is the one with the closest BE value with respect to the reference DLPNO-CCSD(T) value.

We also calculated the BE with the golden standard CCSD(T)/aug-cc-pVTZ on the same ONIOM(B97D3:xTB-GFN2) sample used in the previous test. 
The BE relative errors of B97D3 and the DLPNO-CCSD(T) respect to CCSD(T) are 0.9 and -0.7 kJ/mol respectively. 
These results validate the performance of both B97D3\cite{B97D} and  DLPNO-CCSD(T)\cite{crambina,DLPNO_acc_1}. 

\subsection{\textcolor{black}{Adsorption sites redundancy reduction}}

\textcolor{black}{{During the geometry optimization different NH$_3$ starting points may end up in the same minimum of the potential energy surface (PES), due to the complexity of the PES and the relatively weak interaction energy. For instance, many identical structures differ only on the permutation between the ammonia hydrogen atoms. This redundancy in the adsorption sites was, therefore, reduced by comparing the RMSD and $\Delta$BE between all considered structures and discarding the cases for which  RMSD $\leq 1$ \AA $\,$ and  $|\Delta \mathrm{BE}|$ $\leq$  1 kJ/mol. After cleaning, a total of 77 unique structures from the total 162 starting points were analyzed.}}

\subsection{Machine Learning Binding Energies classification}\label{subsec:ML}

Once the BE distribution\textcolor{black}{{, without sites redundancy,}}  was obtained, a clustering procedure has been performed to collect data. 
Cluster analysis, or clustering, is an unsupervised Machine Learning technique that involves the grouping of data points. 
This grouping is done in such a way that the members of the same cluster can be considered ``\textit{similar}'' in some way (e.g. through metrics like the L2 distance). 
In our case we exploited hierarchical agglomerative clustering (HAC), where an hierarchy of clusters is built with a bottom-up approach: each observation starts in its own cluster, and pairs of clusters are merged as one moves up the hierarchy. 
Sets of observations are linked via the so-called linkage criterion. 
The algorithm will merge the pairs of cluster that minimize this criterion. 
In our study we used the Scikit-Learn's implementation of HAC\cite{scikit-learn}, using the minimum-distance linking criterion (namely ``\textit{single}''), specifying an a-priori number of clusters of 2 (\textit{i.e.} the number of clusters that we want to find). 
Finally, we scaled every feature to [0,1] in order to obtain scaled invariance.

\section{Results and discussion}\label{sec:results}

NH$_3$ usually behaves as a strong hydrogen bond acceptor, due to the  negative electrostatic potential in the nitrogen lone pair region, while being a very weak hydrogen bond donor. 
For instance, the NH$_3$ crystal structure\cite{nh3_crystal} shows only very weak hydrogen bonds between the NH$_3$ molecules, the N$\cdots$H distance being as large as $2.35\,\angstrom$. 
Indeed, our results basically show NH$_3$ acting as a strong H-bond acceptor of the dangling hydrogen of the icy grain and a weak H-bonding donor towards the water oxygen dangling atoms.

After the harmonic frequency analysis, 16 samples show only one imaginary frequency in the  [-50,-8] $\mathrm{cm}^{-1}$ wave-number range. 
Since the imaginary frequencies fall at very low wave-numbers and do not reflect nuclear motion  of the NH$_3$ position, we  kept also these structures to improve the statistics of the BE distribution,
as their very low values do not alter  the final BH(0) values.

\subsection{\textcolor{black}{NH$_3$ desorption rate prefactor}}
\textcolor{black}{{In the desorption process, the desorption rate can be expressed as: $k_{des} = \nu(T) e^{-\frac{BE}{k_B T}}$, where $\nu(T)$ is a pre-exponential factor that takes into account entropic effects, while the enthalpic contribution is inside the exponential part. In order to give reliable data to be used in astrochemical models and/or to have a connection with experiments, a pre-exponential factor must, therefore, be provided together with the BE. Usually, depending on the substrate and adsorbate, a value between $10^{12}-10^{13}$ s$^{-1}$ is assumed in experiments or as a first approximation in modeling studies, as reported by Hasegawa and Herbst \cite{HasegawaHerbst1993} (see e.g. the discussion in Minissale et al. \cite{minissale2022}). We prefer to adopt the transition state theory within the immobile adsorbate approximation \cite{Tait2005,minissale2022} to estimate the prefactor:}}
\begin{equation}\label{eq:apprx_pre_exponential}
    \nu(T) = \frac{\mathrm{k_B T}}{\mathrm{h}} \Biggl(\frac{2 \pi \mathrm{m k_B T}}{\mathrm{h}} \Biggr) \mathrm{A} \frac{\sqrt{\pi}}{\sigma \mathrm{h}^3} \bigl(8 \pi^2 \mathrm{k_B T} \bigr)^{\frac{3}{2}} \sqrt{\mathrm{I_x I_y I_z}} \; ,
\end{equation}
\textcolor{black}{{where $\mathrm{k_B}$ is the Boltzmann constant, m the mass of the molecule, h the Planck constant,  A is the surface area per adsorbed molecules  usually  assumed to be $10^{13} \mathrm{N_a}/$\AA$^2$, $\mathrm{I_i}$ is the \textit{i-esimal} adsorbate principal moment of inertia, and $\sigma$ is the symmetry adsorbate rotation factor.  For  NH$_3$, the principal moments of inertia are 2.76, 1.71, 1.71  a.m.u.$\times$\AA$^2$, $\sigma$=3  and m=17 a.m.u.
When using these values and a desorption peak at T$_{des}$ = 100 K, the pre-exponential factor results $1.94\times10^{15}$ s$^{-1}$ \cite{minissale2022}. This value is recommended in association with the BE values computed with quantum mechanical approaches similar to those described in the present work.}}

\subsection{BE evaluation: calorimetric versus TPD reference}
In the BE calculations, the definition of the ``free'' grain structure, from which the $\mathrm{E}^{iso}_{grn}$ is computed, is crucial and may differ depending on what process one is simulating, while that for the NH$_3$/grain adduct ($\mathrm{E}^{iso}_{ads}$) is unambiguous. 
Usually, when dealing with adsorption on extended surfaces of metal or oxide materials, the reference structure is the bare isolated surface, fully optimized at the given level. 
In such cases, the forces keeping the metal atoms or the ions in place, are much stronger than the BE with the adsorbate and, therefore, the whole structure is little affected by the interaction. 
In the present case, the icy grain is held by forces of the very same nature of those occurring between the adsorbate and the water molecules within the grain. 
Therefore, it may happen that, during the geometry optimization of the adsorbate/grain complex, the grain structure will be altered in such a way that the deformation energy 
$\delta\mathrm{E}^{grn}_{def} =\mathrm{E}^{c}_{grn}-\mathrm{E}^{iso}_{grn}$ 
becomes negative, \textit{i.e.}, the \textit{deformed grain} is more stable than the isolated starting one. 
In other words, the geometry relaxation  induced  by the adsorbate brings the icy cluster in a new local minimum, slightly deeper than the initial one. 
This only happens in a few cases, especially when the Model zone is redefined due to large movements associated to the NH$_3$ molecule. 
To solve this ambiguity in the definition of the deformation energy, we chose, as a starting structure for the isolated cluster to be optimized, the one resulting after the interaction of NH$_3$. 
In this way, $\delta\mathrm{E}^{grn}_{def}$ will be always positive. 
We defined these two approaches  considering different reference pristine grain geometry, as ``calorimetric'' (original initial grain geometry) and ``TPD'' (reference grain geometry after adsorption), respectively. 
The BE distributions from the two approaches will be presented and discussed in the following.

\begin{figure}
\begin{center}
\includegraphics[width=0.8\columnwidth]{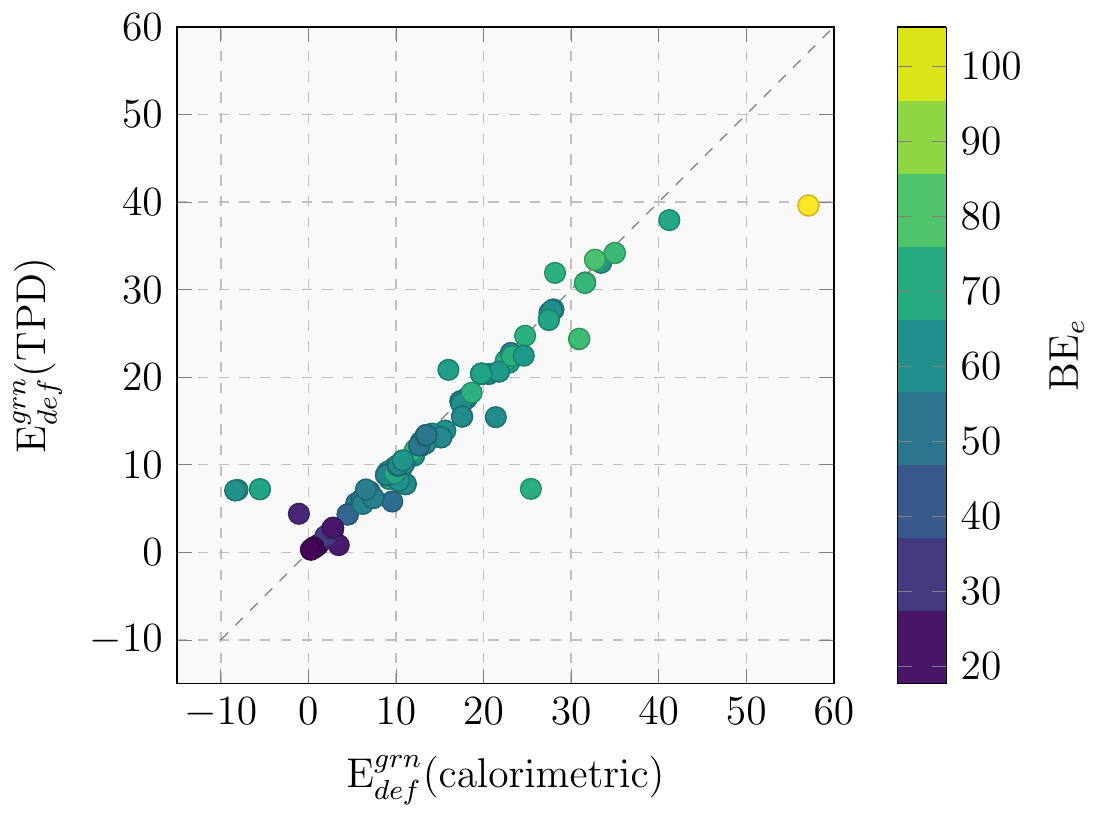}
\end{center}
\caption{Correlation of ``TPD''  \emph{vs} ``calorimetric'' deformation $\mathrm{E}^{grn}_{def}$ energies. The colormap shows the corresponding electronic interaction BE$_e$ (see equation \ref{eq:BE_decompose}) associated to each point. All data in kJ/mol.}
\label{fig:deformation_cal_tpd}
\end{figure}

In the ``calorimetric'' approach, as in microcalorimetric measurements, it is assumed that the reference system is a clean unperturbed surface, and that the heat of adsorption occurs when the adsorbate arrives on the surface from the gas-phase. 
In the temperature programmed desorption (``TPD''),  the molecule is first adsorbed on the surface, and then the temperature is raised  up to the point in which the adsorbate leaves the surface. 
Clearly, when the surface is made by water ice, what is left after desorption cannot be considered equivalent to an unperturbed pristine icy surface, as in the ``calorimetric'' approach.
These two approaches may lead to different BE values, as shown in figure \ref{fig:deformation_cal_tpd}, which correlates the deformation energy $\mathrm{E}^{grn}_{def}$ contribution to the BE computed with both the ``TPD'' and ``calorimetric'' approached. 
The purely electronic BE$_e$ (which is free from the deformation energy) is also put as a reference colored bar. 
As expected, the two approaches lead to the same results for most cases. Nevertheless, there are some exceptions, like some samples with low deformation energy values, in which the surface restructuring leads to a negative deformation energy in the ``calorimetric'' approach. 
The other two outliers ($\mathrm{E}^{grn}_{def}$(calorimetric) $\sim 25$ and 60 kJ/mol) are due to the formation/breaking of some H-bonds at the interface between high- and low-level zones, thus implying a redefinition of the Model zone itself and, therefore, the displacements of many water molecules. 
In the following, we only refer to the ``TPD'' method to compute the final BE distribution.

\subsection{Binding energy distribution}

\begin{figure}
\begin{center}
\includegraphics[width=0.85\columnwidth]{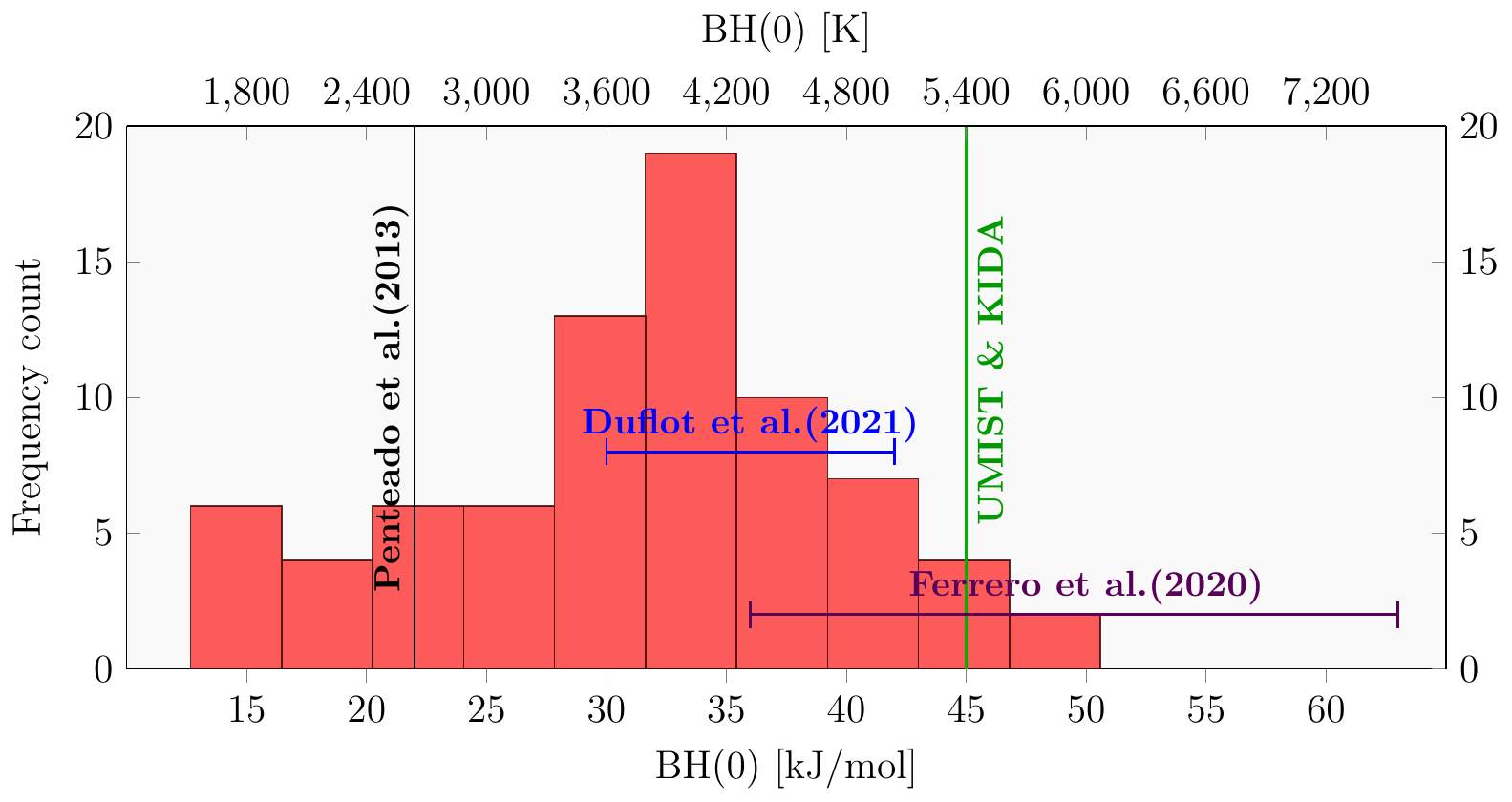}
\end{center}
\caption{BSSE corrected BH(0) distribution at DLPNO-CCSD(T)/aug-cc-pVTZ level. ZPE calculated at ONIOM(B97D3/aug-cc-pVTZ:xTB-GFN2) level.}
\label{fig:be_distribtion}
\end{figure}

\paragraph{New BE distribution versus previous values}
The final BH(0) values (see equation \ref{eq:BE_decompose}) have been organized in a bin width distribution following the Freedman Diaconis estimator\cite{freedman1981histogram}, as shown in figure \ref{fig:be_distribtion}. 
Due to the large number of different adsorbing sites the distribution is asymmetric, with a data dispersion ranging from 12.7 to 50.6 kJ/mol and the mean and mode (the most frequent values) of 31.1 and 33.5 kJ/mol, respectively. 
A fine analysis of the data shows that the deformation energy is the main source of the data dispersion. 
The ZPE plays a minor role in the BH(0), its contribution being of the order of $10\%$ on the total BH(0). 
The ZPE correction decreases the value of the BE by about 10 kJ/mol. 
A value of $\sim$ 45.7 kJ/mol is reported in the astrochemistry databases, which is in the same range, or higher, with respect to the BE values water self-adsorption \cite{wakelam2017binding,mcelroy2013umist}.

The comparison with literature \textcolor{black}{{computational}} BE values by Ferrero et al. (2020)\cite{ferrero2020binding}, computed on a periodic crystalline proton-ordered ice slab model (51.8 kJ/mol) and on an amorphous water slab model (35.9 - 62.8 kJ/mol), is shown in figure \ref{fig:be_distribtion}. 
In that work, the sampling of binding sites on the amorphous slab included just 7 cases and all the interactions found displayed at least NH$_3$ as an acceptor of one hydrogen bond. 
In the work by Duflot et al. (2021),\cite{toubin2021ONIOM}  a procedure similar to the present one  (ONIOM(CBS/DLPNO-CCSD(T):PM6)//ONIOM($\omega$B97X-D/6-31+G$^{**}$:PM6)) was adopted to compute a ZPE corrected BE. 
BE values of 35.9 $\pm$ 11.6 kJ/mol have been computed, in good agreement with our values of 31.1 $\pm$ 8.6 kJ/mol, despite a  very different methodology was adopted to built up the underneath ice. 

\paragraph{Clustering Analysis}

\begin{figure}[!ht]
\centering
\includegraphics[width=0.90\columnwidth]{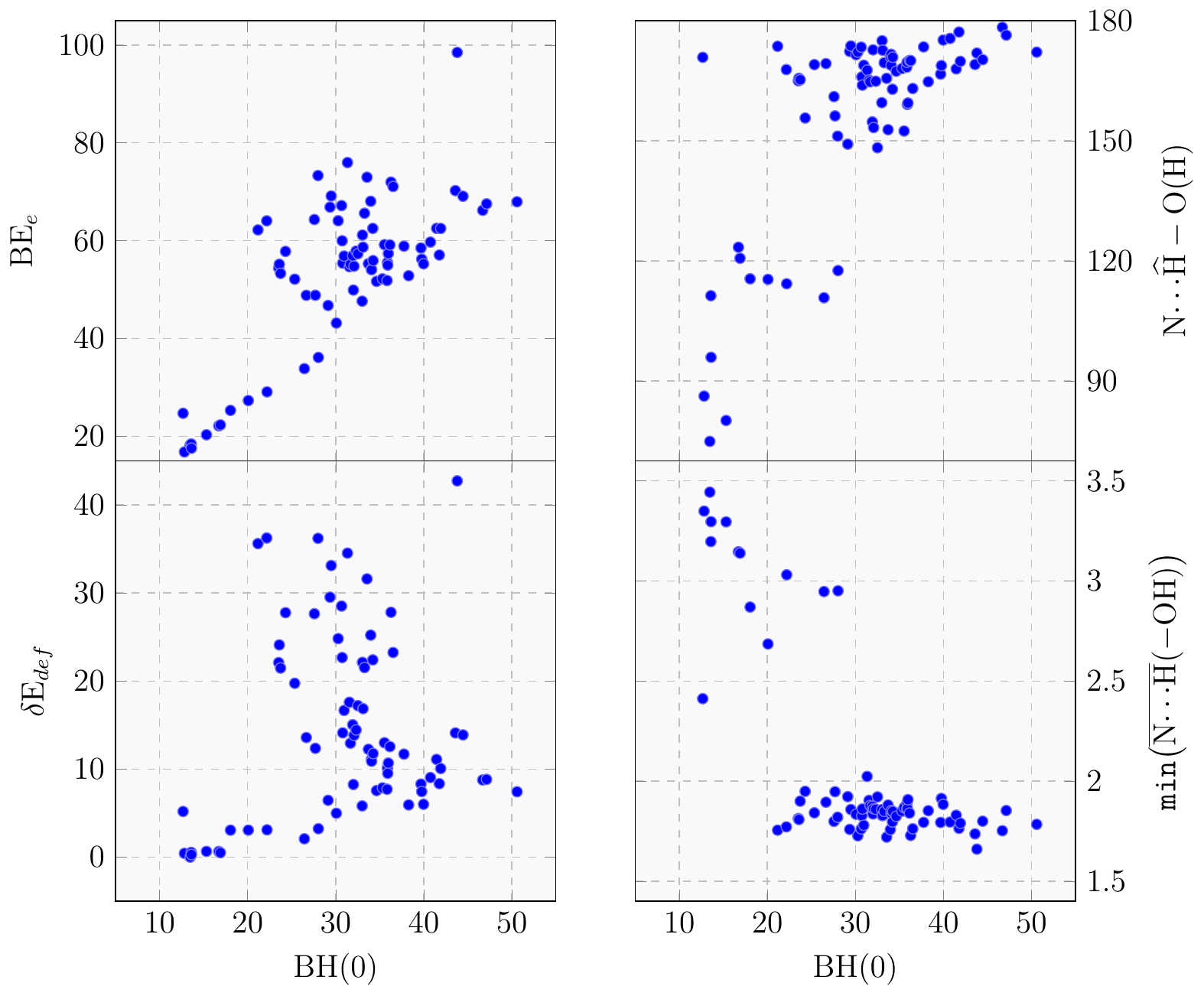}
\caption{Correlation plots between BH(0) and the feature vectors used in the ML clustering. BH(0), BE$_e$ and $\delta\mathrm{E}_{def}$ are units of  kJ/mol. Distances  in $\angstrom\,$ and angles in degrees.  All BH(0) and BE$_e$ are BSSE corrected.}
\label{fig:correlation_features}
\end{figure}

On the final dataset of 77 BEs, a Machine-Learning (ML) based procedure was used in order to correlate the BH(0) with other energetic and geometrical parameters:
\begin{itemize}   
    \item the minimum H-bond distance; \texttt{min}$\bigl(\overline{\mathrm{N}{\cdots}\mathrm{H}}\mathrm{(-OH)}\bigr)$. 
    \item the H-bond angle $\mathrm{N}{\cdots}\mathrm{\widehat{H}-O}\mathrm{(H)}$ referred to the \texttt{min}$\bigl(\overline{\mathrm{N}{\cdots}\mathrm{H}}\mathrm{(-OH)}\bigr)$ H-bond;
    \item deformation energy $\delta\mathrm{E}_{def}$;
    \item the pure electronic BE$_e$.
\end{itemize}
The correlation plots are shown in figure \ref{fig:correlation_features}. 

The plot of both $\overline{\mathrm{N}{\cdots}\mathrm{H}}\mathrm{(-OH)}$ and $\mathrm{N}{\cdots}\mathrm{\widehat{H}-O}\mathrm{(H)}$ revealed a rather clean clustering, in which at high BH(0) values correspond H-bond lengths well below 2 $\angstrom\,$ (NH$_3$ as H-bond acceptor) while at low BH(0)  H-bond distances over 2.5 $\angstrom\,$ (NH$_3$ as H-bond donor). This correlates also with the  $\mathrm{N}{\cdots}\mathrm{\widehat{H}-O}\mathrm{(H)}$ angle, moving from values close to linearity for high BH(0) values, to random values from linearity for the low BH(0) range. Less trivial is the correlation between BH(0) and its different energy components. About the deformation energy $\delta\mathrm{E}_{def}$, a number of points are almost aligned as a baseline in the 0-10 kJ/mol range, while in the region of intermediate BH(0) values the points are quite spread. 
The same erratic trend is seen in the correlation with BE$_e$, revealing that the vast majority of cases exhibit a final BH(0) which is a compromise of a large geometry deformation energy compensated by a large electronic binding energy. 
The few cases at very high BH(0) characterized by small $\delta\mathrm{E}_{def}$ are due to favourable adsorption sites, already suitable to host the NH$_3$ molecule and, therefore, not requiring large structural deformation.

\begin{figure}
\begin{center}
\includegraphics[width=0.85\columnwidth]{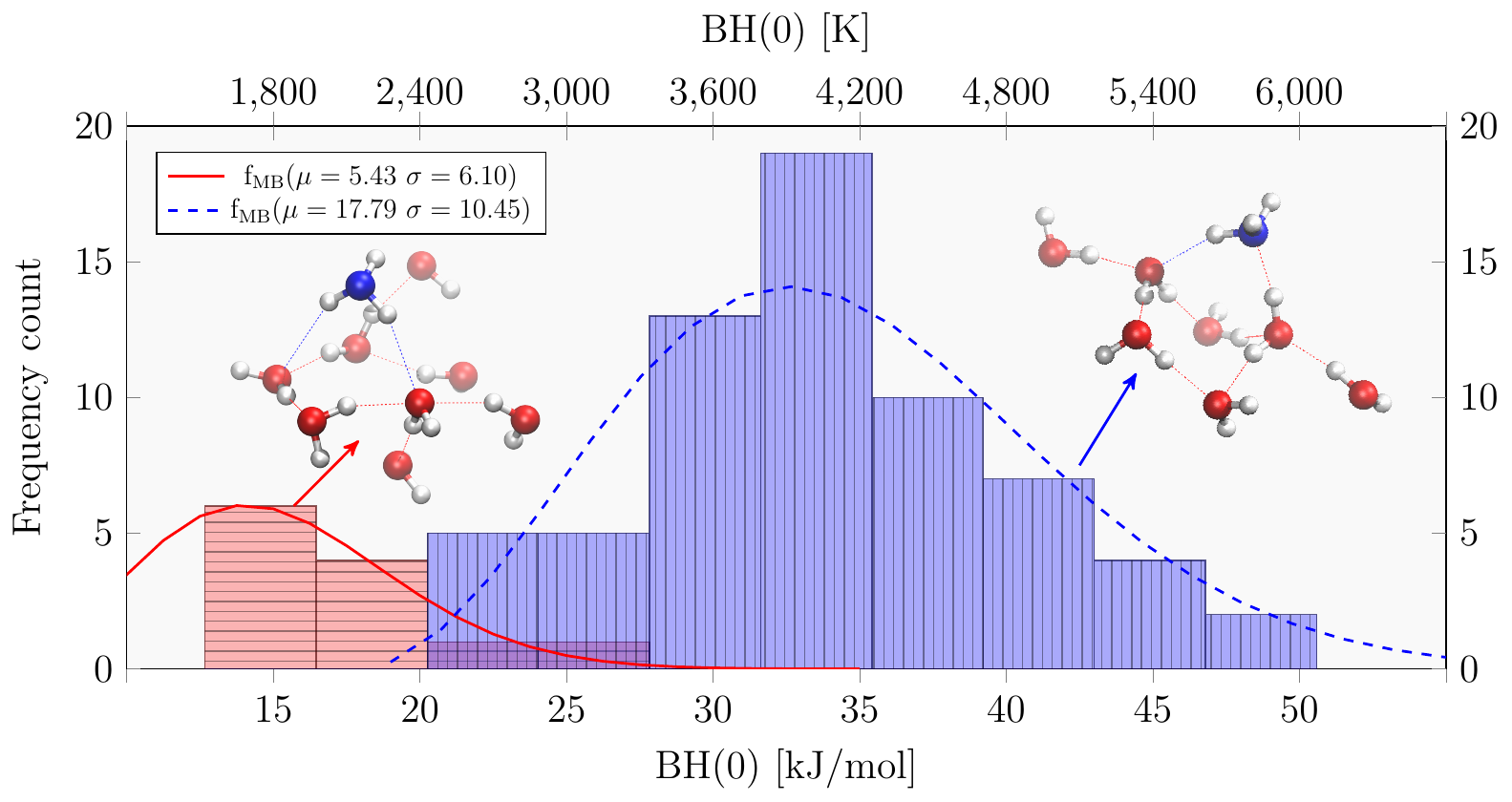}
\end{center}
\caption{ML clustering analysis applied to the BH(0) distribution of figure \ref{fig:be_distribtion}. The continuous red  and dashed blue curves are the f$_\mathrm{MB}$(\texttt{hist}(BH(0)), $\sigma$, $\mu$) Maxell-Boltzmann best fit for the two histogram clusters. The inset shows the Model (high-level) zones of two representative samples, with high (rightmost) and low (leftmost) BH(0) values. Distances  in \angstrom. Atom color legend: oxygen in red, nitrogen in blue, hydrogen in white.}
\label{fig:be_clutering}
\end{figure}

The geometrical clustering analysis applied to the binding energy distribution shown in figure \ref{fig:be_distribtion} is reported in figure \ref{fig:be_clutering}. 
The two clusters rely, as expected by chemical knowledge, on the two possible H-bonds that the ammonia can form with water: the stronger N${\cdots}$H(-OH) and the weaker N-H${\cdots}$O(H$_2$), where the ammonia is respectively H-bond acceptor and H-bond donor. 
In the light of these results, the asymmetric shape of the distribution at low BH(0) is due to the cluster distribution related to the H-bonds in which ammonia is the proton donor.

Moreover, as shown in figure \ref{fig:be_clutering}, the two histogram clusters were fitted with an non-normalized Maxwell–Boltzmann distribution function f$_\mathrm{MB}$(x, $\sigma$, $\mu$):
\begin{equation}\label{eq:MB}
\mathrm{f}_\mathrm{MB}(x, \sigma, \mu)= \frac{(x - \mu)^2}{\sigma^3} \cdot \exp\biggl(- \frac{(x - \mu)^2}{2 \sigma^2}\biggr) \; ,
\end{equation}
where, in our case, \textit{x} are the bins width medium of the BH(0) histogram and $\mu$ and $\sigma$ the distribution parameters.

\pgfplotsset{width=0.85\columnwidth,height=0.35\columnwidth}

\begin{figure}
\begin{center}
\includegraphics[width=0.45\columnwidth]{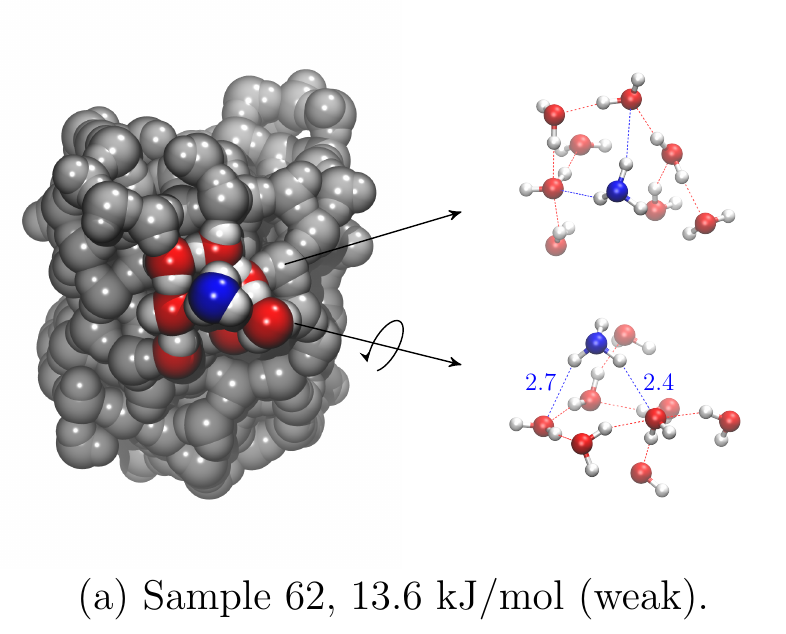}
\hfill
\includegraphics[width=0.45\columnwidth]{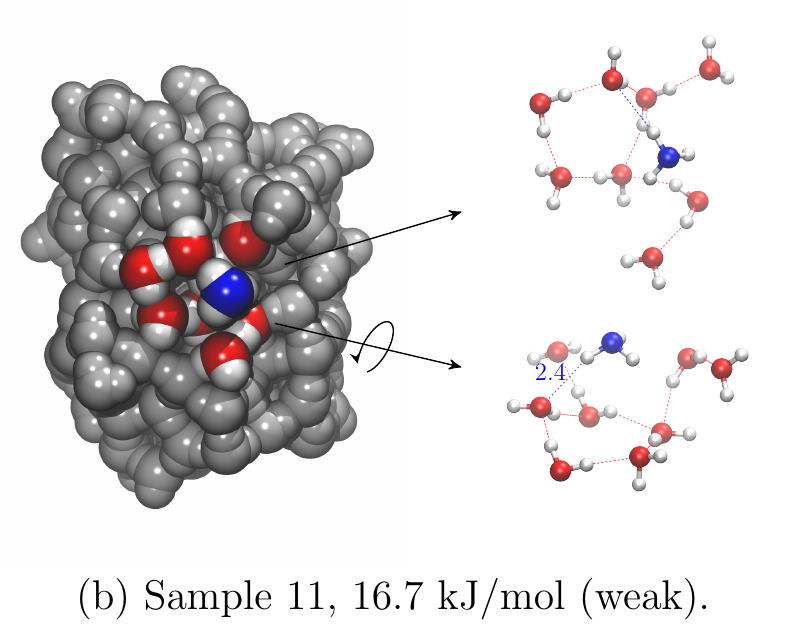}
\\
\includegraphics[width=0.45\columnwidth]{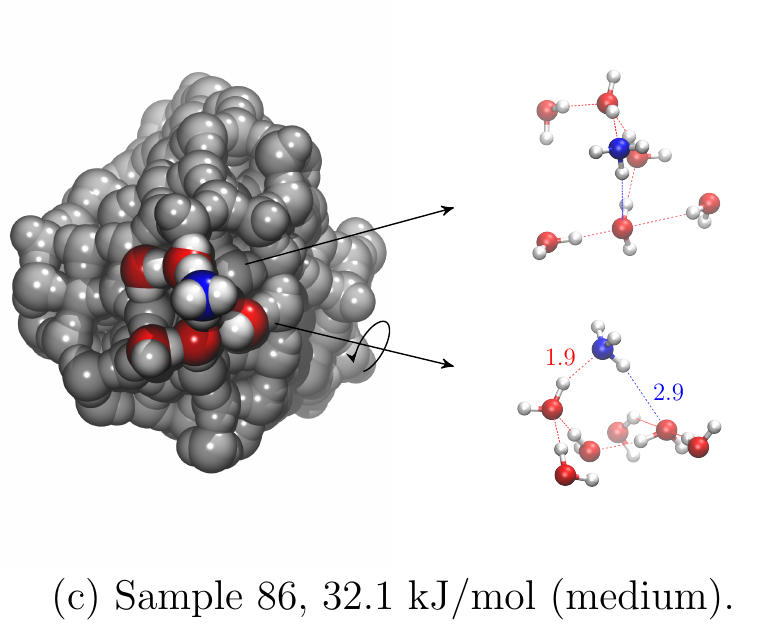}
\hfill
\includegraphics[width=0.45\columnwidth]{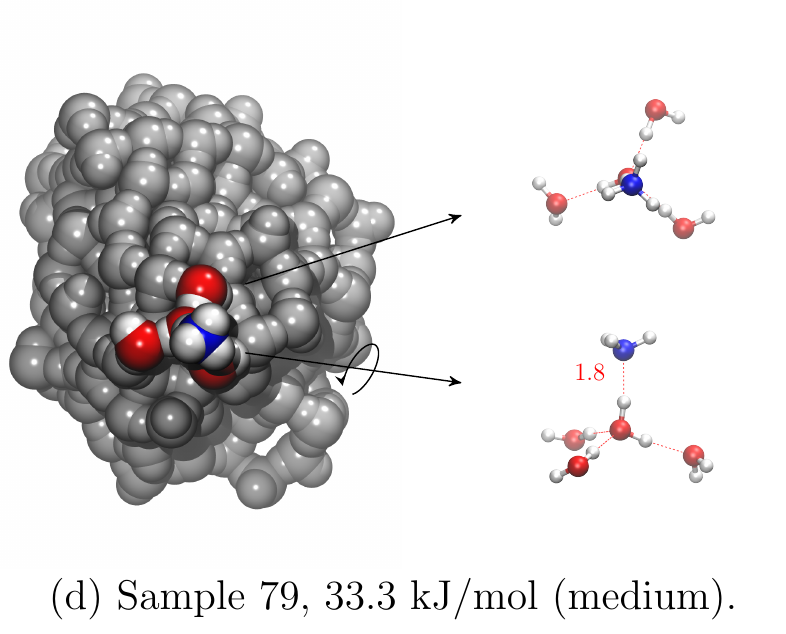}
\\
\includegraphics[width=0.45\columnwidth]{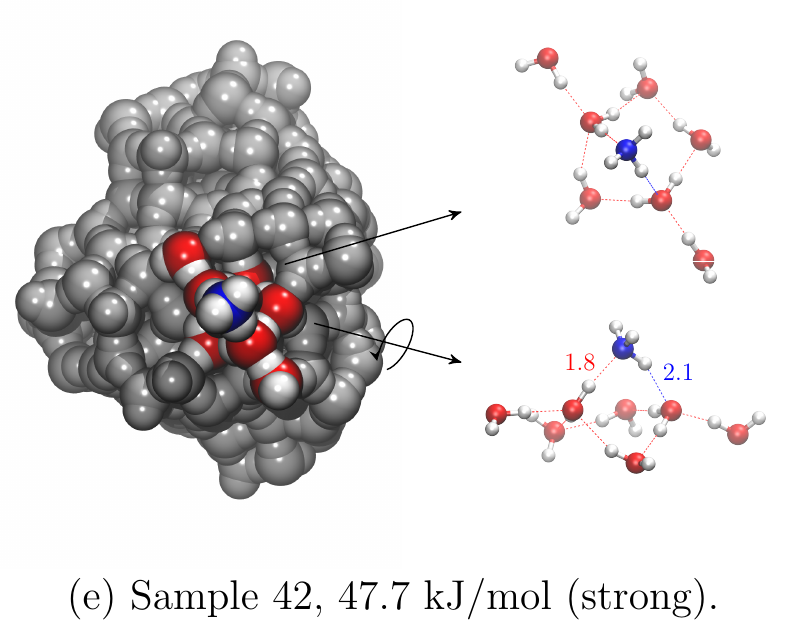}
\hfill
\includegraphics[width=0.45\columnwidth]{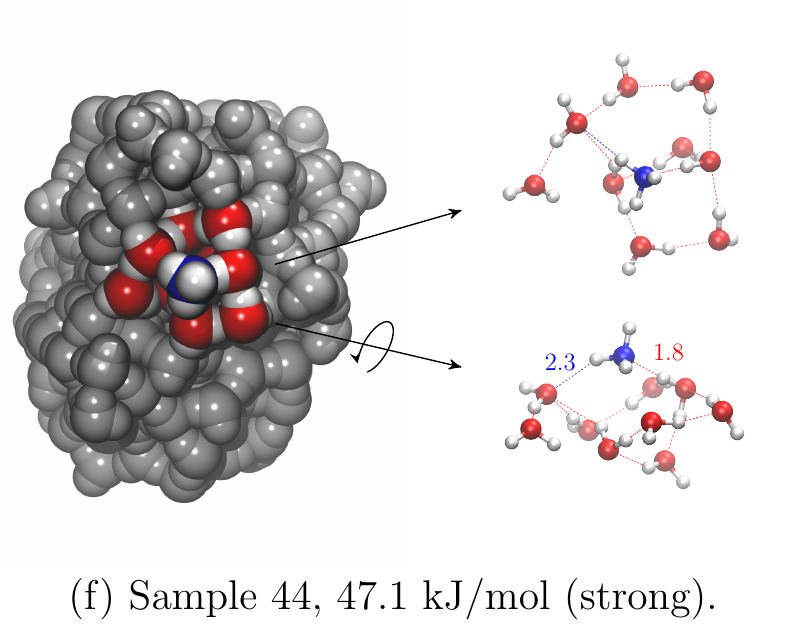}
\\
\end{center}
\caption{Selected cases of weak, medium and strong NH$_3$ BH(0). On the right of each cluster top and later views of the Model zone. Distances  in \angstrom. The online database could be used to easily interact and inspect all the samples, as described in the relative subsection.}
\label{fig:be_strongweak_be}
\end{figure}

Figure \ref{fig:be_strongweak_be} shows a selected number of grain/NH$_3$ structures, spanning from weak to strong values of BH(0), evidencing the already mentioned features of NH$_3$ when interacting through H-bonds.

\textcolor{black}{{Experimental evidence of the tail distribution at very low BE(0) can be searched in the literature, as summarized by Ferrero et al. (2020)\cite{ferrero2020binding}: NH$_3$ TPD experiments on amorphous and crystalline water surfaces are reported by Collings et al.\cite{collings2004laboratory} and He et al. \cite{he2016binding}. However, Collings et al.\cite{collings2004laboratory}, who only carried out experiments on amorphous water ice, did not explicitly derive the NH$_3$ BE. Based on their curve, Penteado et al. (2017)\cite{penteado2017sensitivity} successively estimated a BE equal to 22.5 kJ/mol = 2706 K using the pre-exponential factor equal to $10^{12}$ s$^{-1}$. The BE becomes 3460 K if a pre-exponential factor of $1.94\times 10^{15}$ s$^{-1}$ is used. On the contrary, He et al.\cite{he2016binding} only derived the BE for adsorption on crystalline ice, as they found that NH$_3$ desorbs at the temperature where the amorphous water ice becomes crystalline. Inverting the TPD curve for the crystalline ice adsorption using the pre-exponential factor of $10^{-12}$ s$^{-1}$, He et al.\cite{he2016binding} derived a BE of about  4000 K for a low surface coverage ($\leq 0.5$) and of about 3000 K for a full surface coverage (see Figure 9 of reference\cite{he2016binding}). However, this last value is almost the same as the one derived by the TPD experiments of NH$_3$ adsorbed on gold surface \cite{FrancieleAmmonia,martin2014thermal}, suggesting that a sizable fraction of BE is due to the lateral interactions between NH$_3$ within the adsorbed multilayers and not to the interaction with the ice surface. Moreover the 3000 K BE value (compute with the pre-exponential factor of 10$^{-12}$ s$^{-1}$), becomes 3754 K, with the pre-exponential factor of 1.94$\times10^{15}$ s$^{-1}$, indeed larger than our lower end BE value. Therefore, in both the experimental works presented, the low end of the ammonia BE that we computed was not detected. 
One possibility is that, under low NH$_3$ coverage, NH$_3$ exhibiting very weak BE values (like the one corresponding to our lowest BEs) will easily diffuse to empty sites characterized by higher BE, instead of being entirely desorbed. This process is only effective at moderate NH$_3$ coverage, where sites with high BE values are still availble for occupation.  This indeed happens in TPD experiment in which the thermal heating brings an oversampling of sites at high BE values\cite{minissale2022}. While a detailed astrochemical modeling which may elucidate better this point is postponed to a dedicated work, this discussion also highlights how critical can be the comparison between experimental data extracted from the TPD and the computed one through quantum mechanical calculations if the pre-exponential factor is not treated on the same foot and similar NH$_3$ surface coverage are considered.}}

\subsection{xTB-GFN2 validation}
\begin{figure}
\centering
\includegraphics[width=0.5\columnwidth]{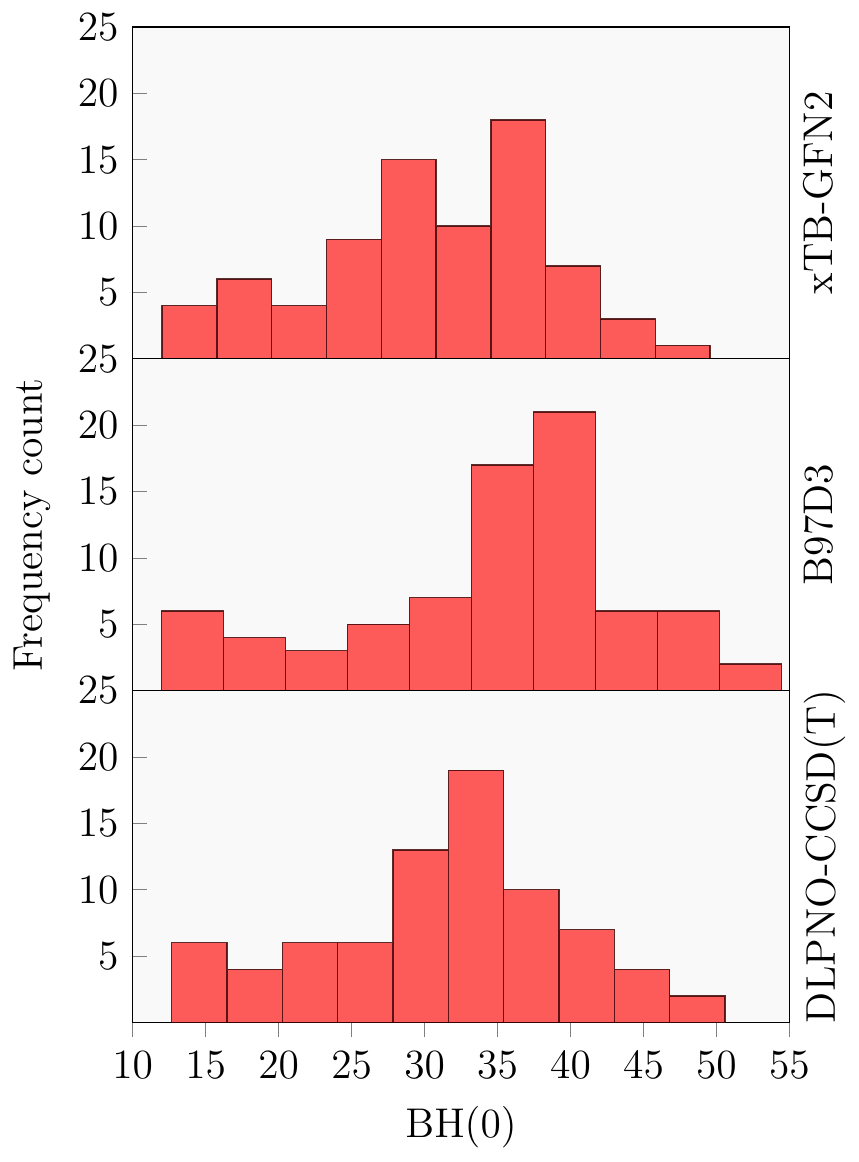}
\hspace{0.05\columnwidth}
\includegraphics[width=0.4\columnwidth]{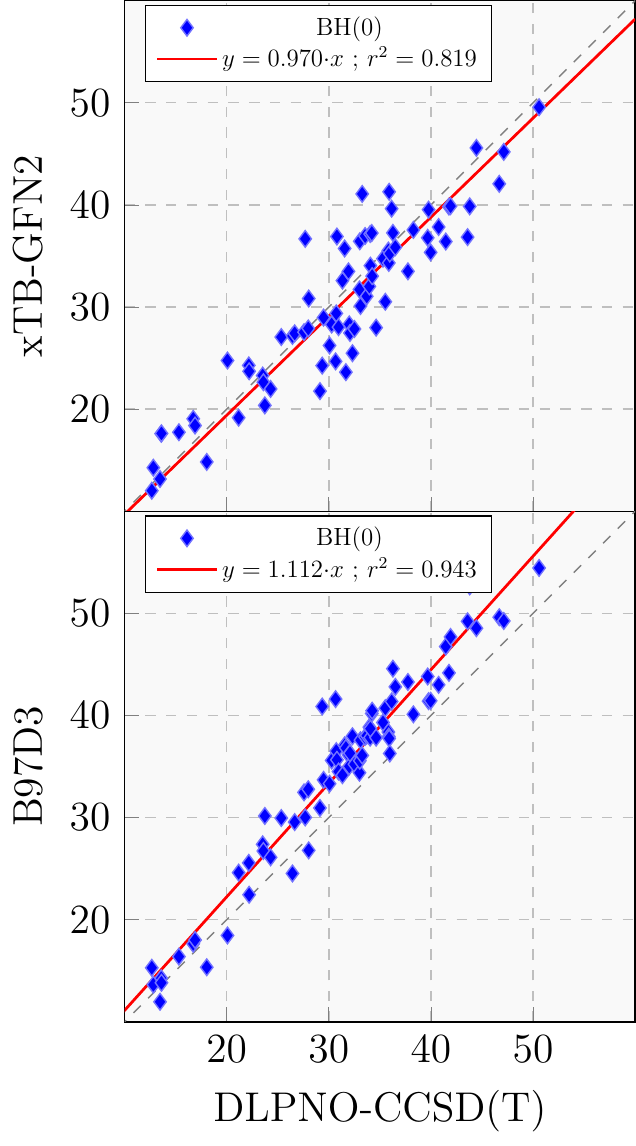}
\caption{Left: BH(0) distributions for DLPNO-CCSD(T), B97D3 and  xTB-GFN2 methods. Right: BH(0) correlation diagrams of B97D3 and xTB-GFN2 against DLPNO-CCSD(T). Each histogram bin width have been calculated with the proper Freedman Diaconis Estimator.
All values in kJ/mol.}
\label{fig:be_distribtions}
\end{figure}

In our recent works,\cite{Aurele_BE_ICCSA,germain2020modeling,Aurele_grain} we adopted xTB-GFN2 as the low-level semi-empirical method. 
The ONIOM procedure requires, to be robust, a low-level of theory giving structures and energies not too far from the high-level one. 
Here, we compare the xTB-GFN2 BH(0) values computed as a single point xTB-GFN2 energy evaluation on the ONIOM optimized geometries, with the more accurate ONIOM ones, computed at DLPNO-CCSD(T) level. 
Figure \ref{fig:be_distribtions} shows the excellent performances of xTB-GFN2, considering its very low computational cost, also in comparison with B97D3 which gives results in better agreement with the DLPNO-CCSD(T) data. 
xTB-GFN2 BH(0) are, instead,  systematically underestimated with respect to the reference. 
The worse GFN2 correlation may be due to the geometric distortion in the Model zone, since it is evaluated at B97D3 level. 

\subsection{Astrochemical implications on NH$_3$ BE distribution}
\begin{figure}
    \centering
    \includegraphics[width=0.7\columnwidth]{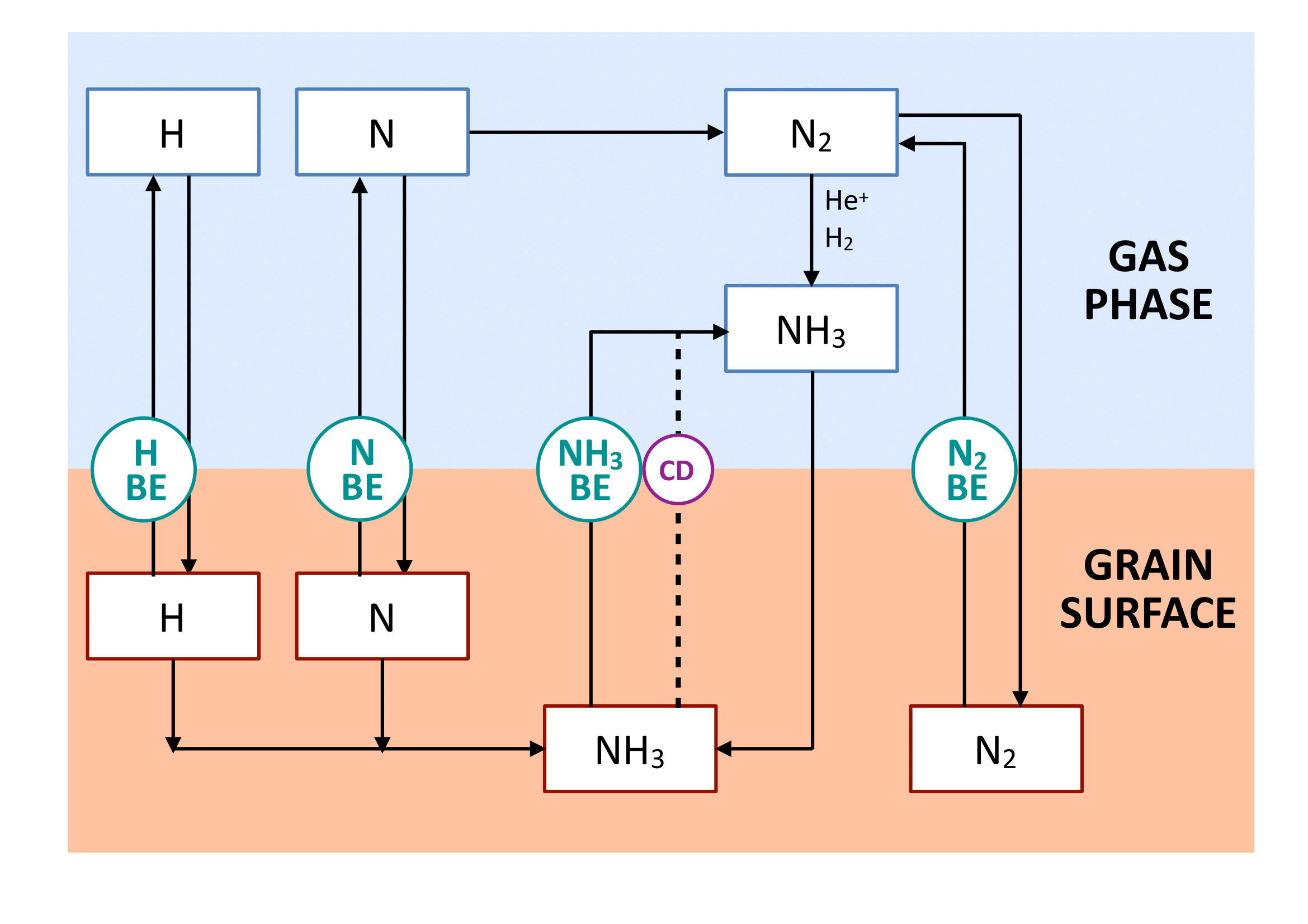}
    \caption{Scheme of the chemistry involving NH$_3$. 
    Ammonia can be synthesised on the grain surfaces because of the hydrogenation of frozen N \cite{Jonusas2020J_nh3grain} (left part of the figure) or in the gas-phase from reactions involving N$_2$ \cite{LeGal2014_nh3gas} and then frozen onto the grain surfaces (right part of the figure).
    Once on the grain surface, NH$_3$ can be thermally desorbed or injected into the gas-phase via the so-called Chemical Desorption (CD) or because of the Cosmic-Ray Desorption (CRD), as marked with a dashed line.
    Both thermal and CRD desorption are governed by the NH$_3$ BE and involve the whole frozen NH$_3$, while CD injects a small fraction ($\leq 1\%$) of the NH$_3$ formed by the N hydrogenation on the grain surface.}
    \label{fig:nh3-chemistry}
\end{figure}

As mentioned in the Introduction, NH$_3$ is ubiquitous in the molecular ISM and can be either gaseous or iced.
Also, NH$_3$ can be formed both in the gas-phase from molecular nitrogen \cite{LeGal2014_nh3gas} and on the grain surfaces by hydrogenation of atomic nitrogen \cite{Jonusas2020J_nh3grain}, as shown in Figure \ref{fig:nh3-chemistry}.
The crucial parameter that governs whether NH$_3$ is in the gaseous or solid form is its BE.
The fact that the NH$_3$ BE is not a single value but a distribution that covers a relative large range of energies, from 1800 to 6000 K (15-50 kJ/mol), can have an important impact (see \textit{e.g.} Grassi et al. \cite{grassi2020novel}).

While gaseous ammonia in warm ($\geq 100$ K) regions does not present any particular puzzle, its presence in cold objects might.
The most extreme example is the gaseous ammonia observed in prestellar objects. 
In L1544, a very well studied prestellar core \cite{caselli1999co}, the dust temperature at the center of the condensation is only 7 K \cite{keto2010dynamics} and ammonia should be completely frozen onto the grain mantles \cite{aikawa2012prestellar,sipila2019does}.
On the contrary, ammonia is observed to be gaseous \cite{crapsi2007observing}.
Various reasons have been proposed, mainly that ammonia is desorbed from the grain mantles because of the chemical energy released by its formation, which is believed to be due to the hydrogenation of N (see \textit{e.g.} Sipila et al. 2019 \cite{sipila2019does}).
These authors found that a bit less of 1\% of the ammonia formed on the grain icy surfaces could be necessary to reproduce the observed values. 
However, these authors also modeled the possibility that the ammonia BE is smaller than the standard high value and considered the cases with BE equal to 1000 and 3000 K (8 and 25 kJ/mol), respectively.
As it could be expected, an ammonia BE equal to 1000 K would result in a too large gaseous ammonia abundance with respect to the observed value.
However, if one considers the BE distribution of figure \ref{fig:be_clutering}, about 3\% of the frozen ammonia would have a BE equal to 1800 K (15 kJ/mol) so that, very likely, the predictions would be in agreement with the observations.

\section{Summary \& Conclusions} \label{sec:conclusions}

In this paper we provide a new framework to compute the binding energy (BE) distribution of any relevant interstellar species adsorbed at the surface of an icy grain mantle, in a reproducible and user friendly automatized way. 
Two main parameters are controlled by the user: the ONIOM high-level zone size, which should be large enough to account for all the H-bond interactions with the ice, and the DFT method for geometry optimization (and subsequent frequency analysis). 
The framework can be divided in four subsequent blocks: 
\begin{enumerate}
    \item Building up of the grain model and choice of the species to be absorbed.
    \item Sampling of all possible binding sites on the icy grain model by an automatic unbiased procedure and geometry optimization with the low level of theory (xTB-GFN2).
    \item Geometry optimization and zero point energy correction using the ONIOM method (B97D3:xTB-GFN2).
    \item Final ONIOM single point (SP) energy refinement with a higher level of theory (DLPNO-CCSD(T)//B97D3:xTB-GFN2).
\end{enumerate}
The first two tasks are encoded in the ACO-FROST program\cite{Aurele_grain} (see also \ref{subsec:ACO_frost}).
An extensive benchmark applied to the ammonia case is reported in the Methodology section, where we demonstrate the performance of the chosen methodology, highlighting its excellent compromise between accuracy and computational cost. 
Moreover, we also demonstrated in a dedicated section that the same distribution calculated at full xTB-GFN2 level is similar to that at ONIOM(DLPNO-CCSD(T)//B97D3:xTB-GFN2), which confirms the robustness of GFN2 despite its cost is orders of magnitude smaller than DFT and DLPNO-CCSD(T).

We highlight a particular aspect which needs to be treated with particular care: the reference of the bare water grain. 
This attention is due to the cooperativity and mobility of the H-bond network, that, when the bare grain is optimized after removing the adsorbate, can lead to strong rearrangements which may result in negative deformation energy (which should be always positive). 
For this reason, we propose and compare two different references for the bare icy surface, which somehow mimic the two experimental techniques used to study such phenomenon: ``TPD'' (each reference is obtained after adsorption, i.e. the NH$_3$ and the bare grain structure re-optimized) or ``Calorimetry'' (the reference is the starting optimized bare grain, before site sampling).

The final ZPE- and BSSE-corrected BE distribution (BH(0)) for ammonia shows, as expected by our 77 unbiased samples, all the possible interactions of NH$_3$ with a water surface, acting as H-bond donor and/or acceptor. 
This variety of BE is made possible by the large number of chemically different binding sites that the built icy grain model presents (\textcolor{black}{{not}} only in terms of dangling species, but also from a morphological point of view of the global structure).
Using an unsupervised Machine Learning clustering technique, we correlate the structures and their BH(0). The two found clusters found with the ML algorithm, can be approximate by two Maxwell–Boltzmann distribution functions with a first peak around 34 kJ/mol (or $\sim 4000$ K) and the second one at $\sim 15$ kJ/mol (or $\sim 1800$ K).
As expected, the asymmetric shape at low BH(0) is due to ammonia acting as H-bond donor, while at high BH(0) we found ammonia acting as both donor and acceptor from a variety of ice dangling hydrogen atoms whose propensity to make H-bonds is modulated by the cooperativity of the H-bond network within the grain. The first peak of the NH$_3$ BH(0) distribution matches very well with the data in the literature, both from experimental and theoretical works. On the contrary, we show for the first time the presence of a second peak at lower BH(0). We discuss how this second peak may explain the long-standing puzzle of the presence of ammonia in cold and dense ISM.

In summary, the major novelty of our work is the development of a framework with a general applicability to simulate all statistical meaningful variety of binding sites of a species adsorbed on icy surface, with high accuracy at reasonable computational cost. 

It allows to produce realistic BE distributions of interstellar molecules, which is a breakthrough with important implications in Astrochemistry.
Our results point toward a more complex scenario about BE than previously thought, as BE in astrochemical models are very often assumed to have a single or very few values, which is an oversimplification of the reality. 

Finally, the presence of low BE has definitively an important impact on our understanding of the chemical evolution of the molecular ISM.

\subsection{Online Database}\label{par:database}
To easily handle the large data set of BE samples (atomic coordinates and BH(0) values), we developed and made public available a web site based on the molecule hyperactive JSmol plugin (Jmol: an open-source Java viewer for chemical structures in 3D\footnote{\href{http://www.jmol.org/}{http://www.jmol.org/}}). The extended electronic version of the calculated results, the 77 optimized structures at ONIOM(B97D3/aug-cc-pVTZ:xTB-GFN2) level, are available at \url{https://tinaccil.github.io/Jmol_BE_NH3_visualization/}.

\begin{acknowledgement}
This project has received funding within the European Union’s Horizon 2020 research and innovation programme from the European Research Council (ERC) for the project “The Dawn of Organic Chemistry” (DOC), grant agreement No 741002, and from the Marie Sk{\l}odowska-Curie for the project ”Astro-Chemical Origins” (ACO), grant agreement No 811312. 
SP, PU acknowledge the Italian Space Agency for co-funding the Life in Space Project (ASI N. 2019-3-U.O). 
Support from the Italian MUR (PRIN 2020, Astrochemistry beyond the second period elements, Prot. 2020AFB3FX) is gratefully acknowledged.
CINES-OCCIGEN HPC is kindly acknowledged for the generous allowance of super-computing time through the A0060810797 project. LT is grateful to Giovanni Bindi for insightful discussions on ML classification and to JL, LM, PtF and finally to the \LaTeX $\,$ community for the insights on TikZ and PGFPlots packages. 
Finally, we wish to acknowledge the extremely useful discussions with Prof. Gretobape.
\end{acknowledgement}

\begin{suppinfo}

The \texttt{Dataframe.csv} is a table presenting all the Binding Energy (BE) information for each sample, all the values are referred to the "TPD" approach (see the discussion in the paper).

All the computed structures (\texttt{.xyz} files), at ONIOM(B97D3/aug-cc-pVTZ:xTB-GFN2), are available in the \texttt{structures} folder. In this folder are present the complex structures (grain\_molX.xyz) and the Model zone (CP\_grain\_molX.xyz), the number inside (X) the file name is referred to the sample ("n\_sample", in the Dataframe).
The bare grain and ammonia are also prensent (bare\_grain.xyz and nh3.xyz), optimized respectivetly at xTB-GFN2 and B97D3/aug-cc-pVTZ.

Input examples for all the programs are inside the \texttt{example\_input} folder. Since the xTB program is call by "external" keyword in Gaussian16, is also provided the script to interface (i.e. \texttt{xtb-gaussian.sh}) xtb to Gaussian16.

\end{suppinfo}

\bibliography{achemso-demo}

\end{document}